\documentclass[10pt]{article}

\usepackage{amsmath}
\usepackage{placeins}
\usepackage{amssymb}
\usepackage{amsthm}

\usepackage{graphicx}

\usepackage{cite}
\usepackage{color} 
 \usepackage[colorlinks,breaklinks]{hyperref}
 \hypersetup{hypertexnames=true,
   linkcolor=blue,anchorcolor=black,citecolor=blue,urlcolor=blue
 }

\DeclareGraphicsExtensions{.png,.pdf}
\graphicspath{{imglocal/}{img/}{./}}

\usepackage[labelfont=bf,labelsep=period,justification=raggedright]{caption}
\usepackage{subcaption}

\makeatletter
\renewcommand{\@biblabel}[1]{\quad#1.}
\makeatother

\newcommand{\giacomo}[1]{}

\date{}

\newtheorem{prop}{proposition}

\begin{document}

\begin{flushleft}
{\Large
\textbf{Rhythmic inhibition allows neural networks to search for maximally consistent states}
}
\\
Hesham Mostafa$^*$, 
Lorenz K. M\"uller, 
Giacomo Indiveri.
\\
Institute for Neuroinformatics, University of Zurich and ETH Zurich, Switzerland \\
E-mail: \{hesham,lorenz,giacomo\}@ini.uzh.ch
\end{flushleft}
\section*{Abstract}
Gamma-band rhythmic inhibition is a ubiquitous phenomenon in neural circuits yet its computational role still remains elusive. We show that a model of Gamma-band rhythmic inhibition allows networks of coupled cortical circuit motifs to search for network configurations that best reconcile external inputs with an internal consistency model encoded in the network connectivity. We show that Hebbian plasticity allows the networks to learn the consistency model by example. The search dynamics driven by rhythmic inhibition enable the described networks to solve difficult constraint satisfaction problems without making assumptions about the form of stochastic fluctuations in the network. We show that the search dynamics are well approximated by a stochastic sampling process. We use the described networks to reproduce perceptual multi-stability phenomena with switching times that are a good match to experimental data and show that they provide a general neural framework which can be used to model other 'perceptual inference' phenomena.

\section*{Introduction}
\label{sec:introduction}

Gamma oscillations are rhythmic patterns of neural activity in the 30-80\,Hz frequency range that have been measured in the extracellular fields of multiple brain areas across many species~\cite{Buzsaki_Wang12}. They have been associated with attention~\cite{Womelsdorf_etal07}, working memory\cite{Deco_Rolls03}, and the execution of cognitive tasks~\cite{Tallon-Baudry_etal97}. Local rhythmic inhibition is a fundamental feature of Gamma oscillations~\cite{Stein_Sarnthein00,buzsaki_draguhn04} that acts to modulate the firing rate of the local circuit as well as its sensitivity to external input~\cite{Cardin_etal09}. Although many studies have elucidated several aspects of the biophysical mechanisms underlying Gamma-band rhythmic inhibition~\cite{Jadi_Sejnowski14,Brunel_Wang03},  its computational and functional role is still a matter of debate~\cite{Buzsaki_Wang12,Jadi_Sejnowski14}.

In this paper, we investigate the functional role of local Gamma-band rhythmic inhibition in rate-based networks of neurons configured as coupled Winner-Take-All (WTA) circuits. The WTA circuits are used as models of local cortical circuit motifs~\cite{Douglas_Martin04}. The strength of the effective connectivity between a pair of coupled WTAs is continuously modulated by the phase difference between their local rhythms. Higher coherence between these rhythms leads to more reliable communication between the WTA pair in line with the ``communication through coherence'' hypothesis~\cite{Bosman_etal12,Fries05}. On a global level, we show that oscillatory inhibition drastically alters the dynamics of coupled WTA networks and allows them to search for activity states that best satisfy the constraints encoded in the network connectivity while being consistent with the externally applied inputs. We show that rhythmic inhibition effectively allows neural networks to solve constraint satisfaction problems which are among the most difficult classes of computational problems. This result is particularly relevant to neural models of sensory processing. A long theoretical tradition casts sensory processing as being a process of inferring the maximally consistent interpretations of imperfect sensory input where consistency is judged according to an internal model of the environment constructed from prior experience ~\cite{von-Helmholtz_Southall25,Friston03,Berkes_etal11}. There are two fundamental questions that arise when considering the neural substrate underlying the search for consistent interpretations: (1) how to \emph{learn and represent} the consistency model; (2) how to use the consistency model to obtain \emph{plausible interpretations} of ambiguous inputs. The networks we describe address these two questions within a biologically realistic framework.

Several neural architectures have been proposed whose dynamics execute a search for configurations that maximally satisfy a set of constraints. One such architecture uses Hopfield networks that are engineered so that their energy functions have the deepest minimum at the maximally consistent state~\cite{hopfield_tank85,Hopfield08} which thus becomes a  stable point of the network dynamics. In general, however, there are other non-optimal network states that are also dynamically stable states. There is no guarantee that the network will not get stuck in these sub-optimal states~\cite{Hopfield08}. Moreover, in the case of ambiguous input, the network does not explore all the best, equally consistent states or input interpretations. A network model for solving constraint satisfaction problems using multi-stable WTA-based oscillators has been described in~\cite{Mostafa_etal13b}. In this model as well, there is no guarantee that the network will not get stuck at locally optimal solutions and the model does not address how the internal consistency model can be learned. Alternative architectures sidestep the issue of getting stuck at locally optimal states or input interpretations by making use of stochastic networks that continuously explore the state space~\cite{Buesing_etal11,Pecevski_etal11}. Such networks can be configured to visit more consistent states with higher probability. However, for an unambiguous input, such networks can not search for, and stabilize at, the fully consistent state,  but would only transiently visit this optimum state~\cite{habenschuss_etal13}.

We show that the issues inherent in these architectures can be avoided by using rhythmic inhibition to drive the search for maximally consistent input interpretations. The networks we describe never get stuck at non-optimal input interpretations; if a fully consistent interpretation of an unambiguous input exists, the network will find and stabilize at that interpretation; if the input is ambiguous or the set of constraints irreconcilable, the network will continuously switch between the most consistent states or input interpretations. In the latter case, we show that the network trajectory can be well approximated by a stochastic Markov chain Monte Carlo (MCMC) sampler which is remarkable considering that the network trajectory is deterministic. We use this continuous switching between equally consistent interpretations to model perceptual multi-stability phenomena. Unlike the majority of previous models though, we start with a 'naive' network and allow it to learn by example, through Hebbian plasticity in the  network connections, what constitutes consistent inputs. The multi-stability phenomenon emerges when the network receives ambiguous input that does not admit a consistent interpretation according to the previously seen examples. The network architecture we describe can be used as a biologically motivated neural substrate on which to ground various ``perception as inference'' theories.

\section*{Results}
\label{sec:results}
\subsection*{Modeling assumptions}
The networks we describe in this paper are composed of a number of local neural circuits that interact through long range excitatory connections and that each receive local oscillatory inhibition. The properties of the oscillatory inhibition model the salient features of the rhythmic inhibition that underlies Gamma oscillations. The following are biological justifications for the various modeling assumptions we use:
\begin{enumerate}
\item \emph{WTA circuits are local neural circuit motifs}: WTA circuits are potential cortical circuit motifs~\cite{Douglas_Martin04}. The WTA circuits can be replaced by any neural circuit as long as this circuit displays a number of distinct firing patterns based on external input and a memory of past firing patterns. Some distinct part of each firing pattern should be characterized by a high enough firing rate so that it can influence the firing pattern in another neural circuit when the two are coupled. 
\item \emph{Oscillatory inhibition is local to each WTA}: Gamma oscillations typically have a local origin~\cite{Buzsaki_Wang12,Stein_Sarnthein00,ainsworth_etal11}. Gamma oscillations in local field potential typically arise from the rhythmic firing of basket cells that have predominantly local arborization~\cite{Fries09}. It is a general phenomenon that faster rhythms tend to develop locally~\cite{buzsaki_draguhn04}.  
\item \emph{Local oscillatory inhibition is strong enough to shut down the activity in the local circuit}: There is strong evidence for the involvement of interneurons containing the calcium binding protein Parvalbumin in the oscillatory discharge underlying Gamma oscillations ~\cite{Sohal_etal09,Cardin_etal09}. Interneurons expressing Parvalbumin, such as basket cells and chandelier cells, mainly target the soma, the axon initial segment, and the proximal dendrites of the excitatory principal cells~\cite{Markram_etal04} making them particularly effective in rhythmically silencing their target neurons.  
\item \emph{The local oscillatory inhibition waveforms have different frequencies}: There is evidence that Gamma oscillations recorded from even nearby regions in the same cortical area have significantly different frequencies~\cite{Ray_Maunsell10}. Phase relations between Gamma oscillations recorded from nearby points in the cortex are continuously changing~\cite{Womelsdorf_etal07,Fries09}, and the phases of local cortical Gamma rhythms vary in an irregular manner~\cite{Burns_etal10}. Different oscillation frequencies is one simple way to obtain continuously changing phase relations. The different local rhythms can be highly coherent and we investigate the effect of this coherence on the model behavior. The only requirement of our model is that the space of possible phase relations is continuously explored with no perfect and persistent phase lock between any of the local rhythms. 
\end{enumerate}
\subsection*{Network description}
\label{sec:gamma-oscill-winn}
\begin{figure}
  \centering
   \begin{subfigure}[b]{0.8\textwidth}
     \includegraphics[width=\textwidth]{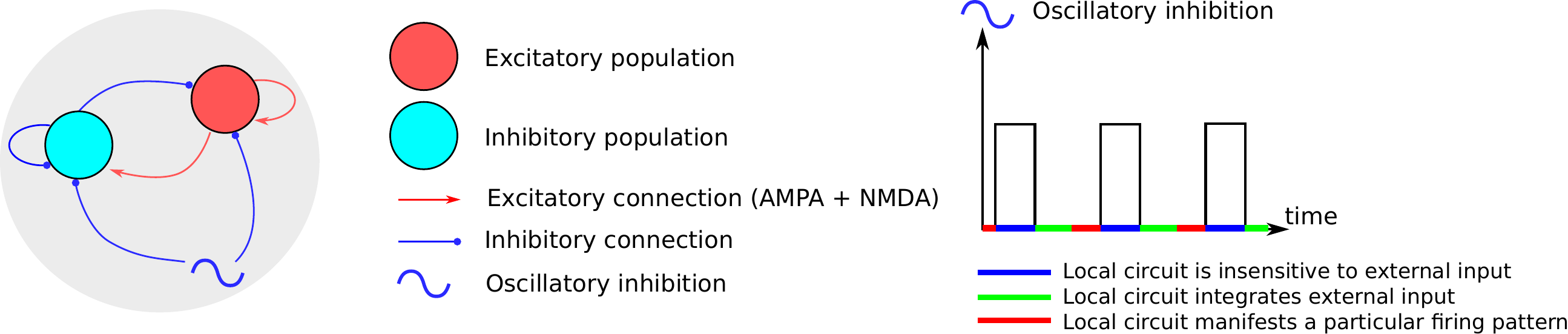} 
     \subcaption{} 
     \label{fig:arch_desc_a}
   \end{subfigure}
   \\
   \begin{subfigure}[m]{0.4\textwidth}
     \includegraphics[width=\textwidth]{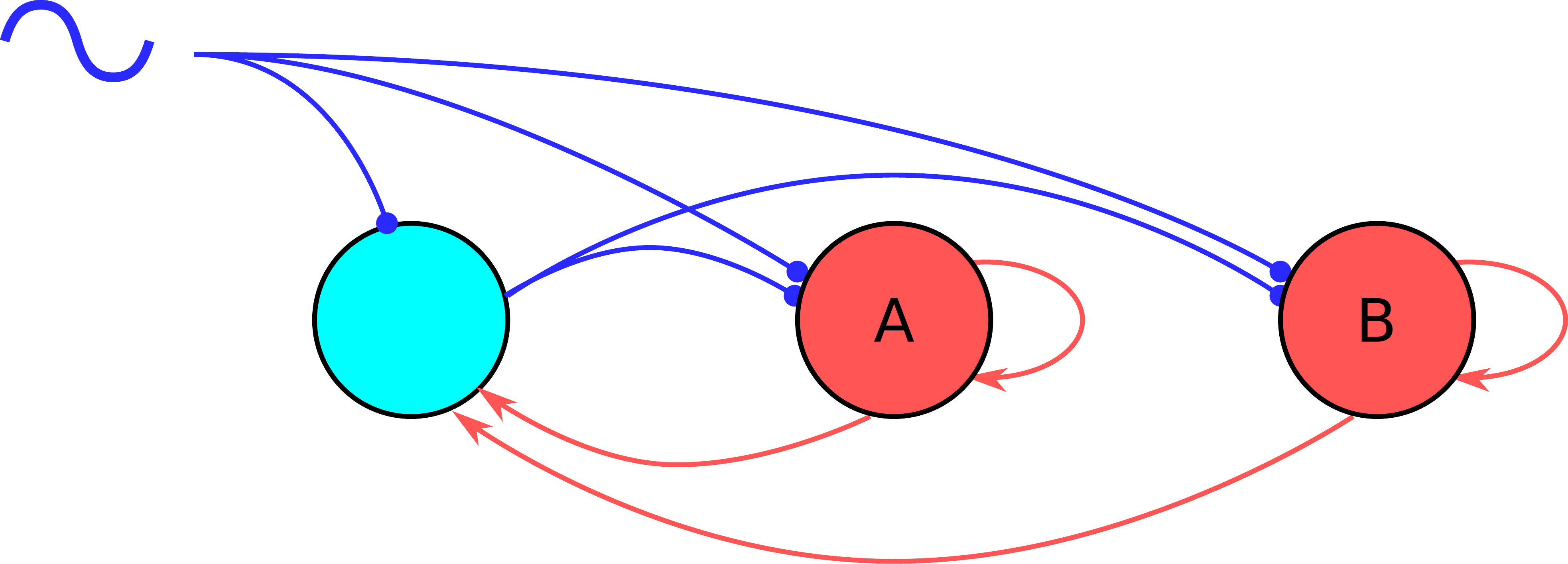} 
     \subcaption{}
     \label{fig:arch_desc_b}
   \end{subfigure}
   \quad \quad
   \begin{subfigure}[m]{0.475\textwidth}
     \includegraphics[width=\textwidth]{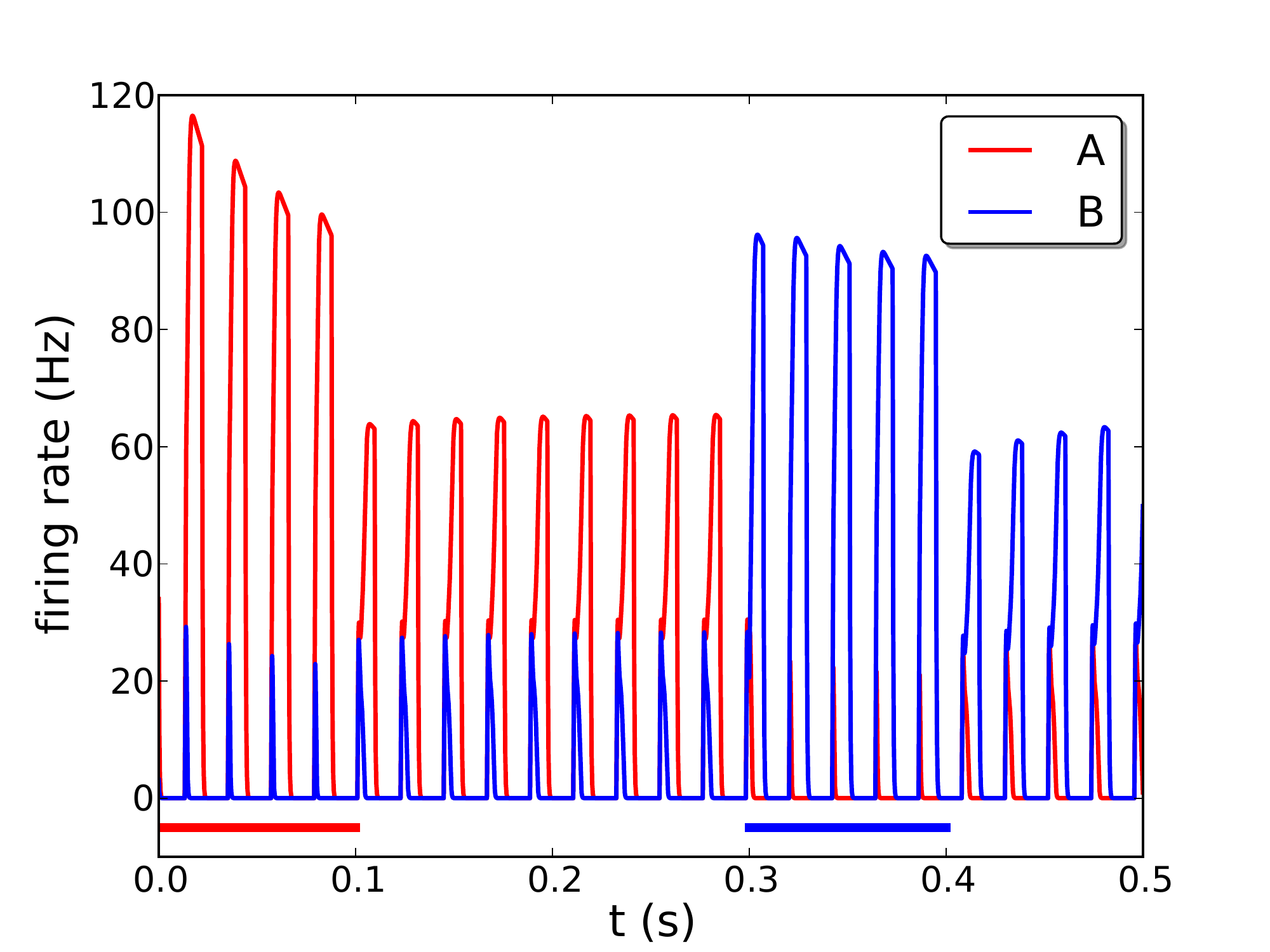} 
     \subcaption{}
     \label{fig:arch_desc_c}
   \end{subfigure}
   \\
   \begin{subfigure}[m]{0.4\textwidth}
     \includegraphics[width=\textwidth]{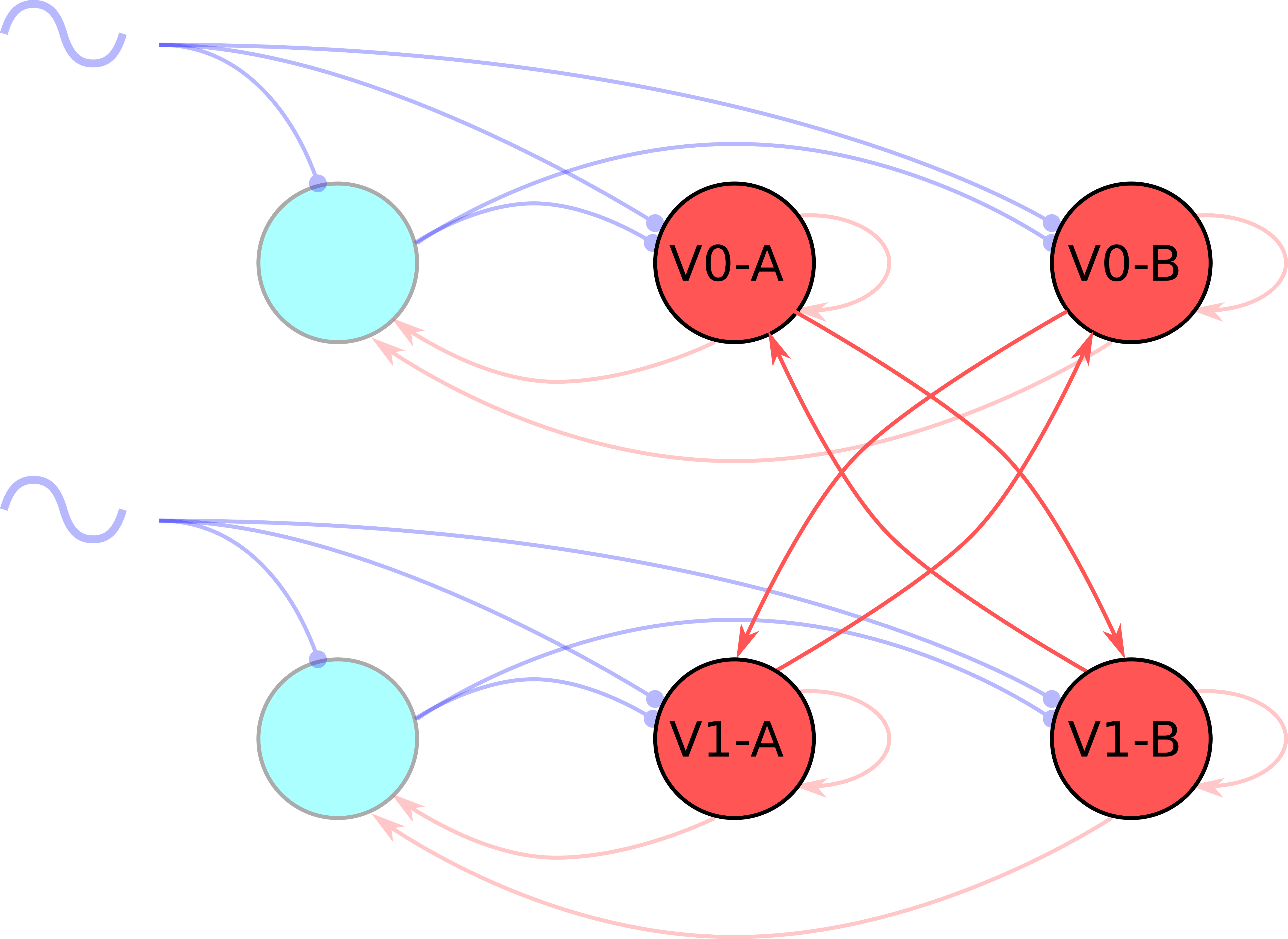} 
     \subcaption{}
     \label{fig:arch_desc_d}
   \end{subfigure}
   \hfill
   \begin{subfigure}[m]{0.5\textwidth}
     \includegraphics[width=\textwidth]{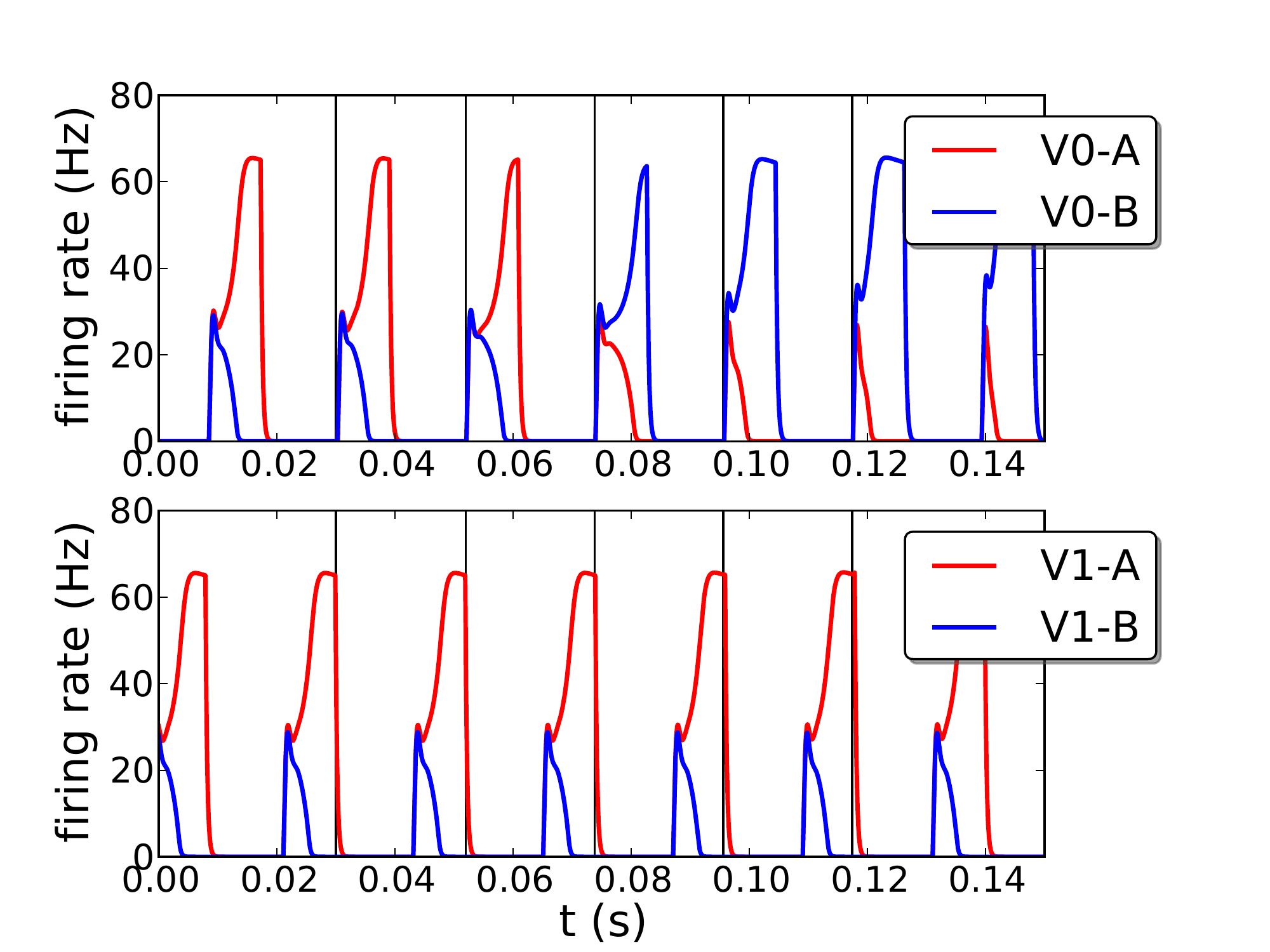} 
     \subcaption{}
     \label{fig:arch_desc_e}
   \end{subfigure}
   \caption{(\subref{fig:arch_desc_a}) Generic structure of a
     local circuit that undergoes oscillatory
     inhibition. Oscillatory inhibition is modeled as a periodic rectangular function. (\subref{fig:arch_desc_b}) A binary WTA circuit
     modulated by oscillatory inhibition. (\subref{fig:arch_desc_c}) Simulation results of
     the binary WTA. Oscillatory inhibition periodically
     shuts down all activity.  External input (colored horizontal bars) can
     bias the winner selection process and in its absence, the identity of the winning excitatory population is
     maintained across the inhibition
     cycles. (\subref{fig:arch_desc_d}) Coupling two WTA circuits
     , \textsf{V0} and \textsf{V1}, to
     enforce the consistency condition
     \textsf{V0}$\ne$\textsf{V1}. (\subref{fig:arch_desc_e})
     Simulation results of the network in
     (\subref{fig:arch_desc_d}) when the frequency of
     the oscillatory inhibition is slightly different in the two WTA
     circuits. Vertical lines are visual guides to
     the end of the high phase of oscillatory inhibition in
     the \textsf{V0} WTA circuit.}
\label{fig:arch_desc}
\end{figure}

We simulate and analyze rate-based population-level networks. Each neural population in a WTA is modeled as a linear threshold unit (LTU) following a mean-field approach~\cite{Amit_Brunel97,Knight00,Mattia_Del-Giudice02} (see Methods section). All excitatory connections have a slow and fast component to capture the effects of the different  time constants of the synaptic currents mediated by slow $NMDA$ receptors and fast $AMPA$ receptors. The network architecture and example simulation results are shown in Fig.~\ref{fig:arch_desc}. 
When a WTA is released from oscillatory inhibition, the excitatory population receiving the largest external input wins and suppresses the activity in the other excitatory populations in the WTA. The $NMDA$-mediated recurrent synaptic current has a time constant that is longer than the oscillatory inhibition period. Since this current is always larger in the winning population, it is able to bias the competition so that the winning population keeps on winning in subsequent inhibition cycles after the external input is removed as shown in Fig.~\ref{fig:arch_desc_c}. 

Figure~\ref{fig:arch_desc_d} shows two coupled WTA circuits in which the coupling connections encode a consistency condition.
Each WTA circuit represents a discrete variable with two states: \textsf{A} and \textsf{B} and the coupling connections dictate that when population \textsf{A} is winning in one WTA circuit then population \textsf{B} should be winning in the other. The  activity of one WTA circuit, however, is able to influence the state of the other WTA only when its interval of heightened activity coincides with the winner selection interval (the interval just after oscillatory inhibition goes low) in the other WTA circuit. This effect is illustrated in Fig.~\ref{fig:arch_desc_e} where we choose slightly different frequencies for the oscillatory inhibition in the two WTA circuits. In the supplementary material, we describe how arbitrary consistency conditions involving more than two WTAs can be realized.

This example shows how the effective strength of the inter-WTA coupling connections depends on the phase difference between their rhythmic inhibition cycles. This is compatible with the hypothesis that the effective strength of the fixed anatomical connections in the brain is modulated by the dynamical state of the communicating neural circuits which allows the quick formation and removal of effective ``communication'' links~\cite{Friston94,Massimini_etal05}. The phase difference between the rhythms of two neural groups is a potential marker for the strength of the effective link between them~\cite{Womelsdorf_etal07,Bosman_etal12}. This view is particularly appealing in the case of gamma oscillations due its rhythmic modulation of excitability and firing rate~\cite{Fries05}.

\subsection*{Properties of the network trajectory and the search for consistent states}
A WTA circuit can be involved in a number of different consistency conditions. As the rhythmic inhibition wears off in one WTA circuit, other WTA circuits that are coupled to it can influence its state so as to satisfy the consistency conditions encoded in the pattern of coupling connections (see Fig.~\ref{fig:arch_desc_d} for example). Exactly which consistency condition, if any, gets satisfied by the new state/winner of the WTA circuit depends on the level of activity in the WTA circuits connected to it, which in turn depends on the phase of the inhibition cycle in these WTA circuits. Since the phase relations between the WTA circuits are continuously changing, the relative strengths of the different consistency conditions are also continuously changing.

If we consider each WTA as a discrete-valued variable whose value is the identity of the last winning population, we can enumerate all possible configurations of a network of interacting oscillatory WTA circuits; for example, a network of $N$ coupled WTA circuits with two excitatory populations each can have $2^N$ possible configurations. Define $d(t)$ as the discrete-valued, continuous-time, dynamical variable that denotes the configuration of the network at time $t$. This variable is updated each time the local oscillatory inhibition shuts down the activity in a WTA circuit to reflect the identity of the WTA circuit's last winning population at that time. We can prove the following two results about the trajectory of $d(t)$:

\begin{prop}
The only fixed points of $d(t)$ are the configurations that satisfy all consistency conditions
\label{prop:fixedpoint}
\end{prop}

\begin{prop}
\label{prop:aperiodic}
$d(t)$ is not periodic as long as it has not reached a fixed point
\end{prop}

\begin{figure}
   \begin{subfigure}[b]{0.4\textwidth}
     \includegraphics[width=\textwidth]{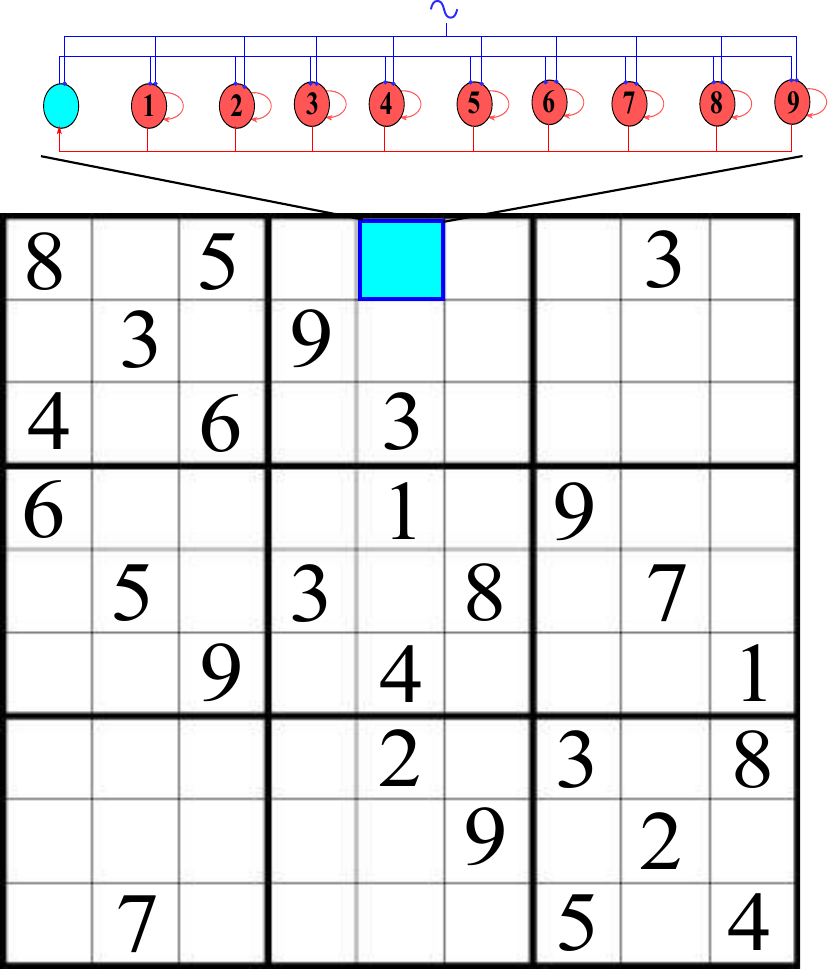} 
     \subcaption{} 
     \label{fig:sud_a}
   \end{subfigure}
   \hfill
   \begin{subfigure}[b]{0.6\textwidth}
     \includegraphics[width=\textwidth]{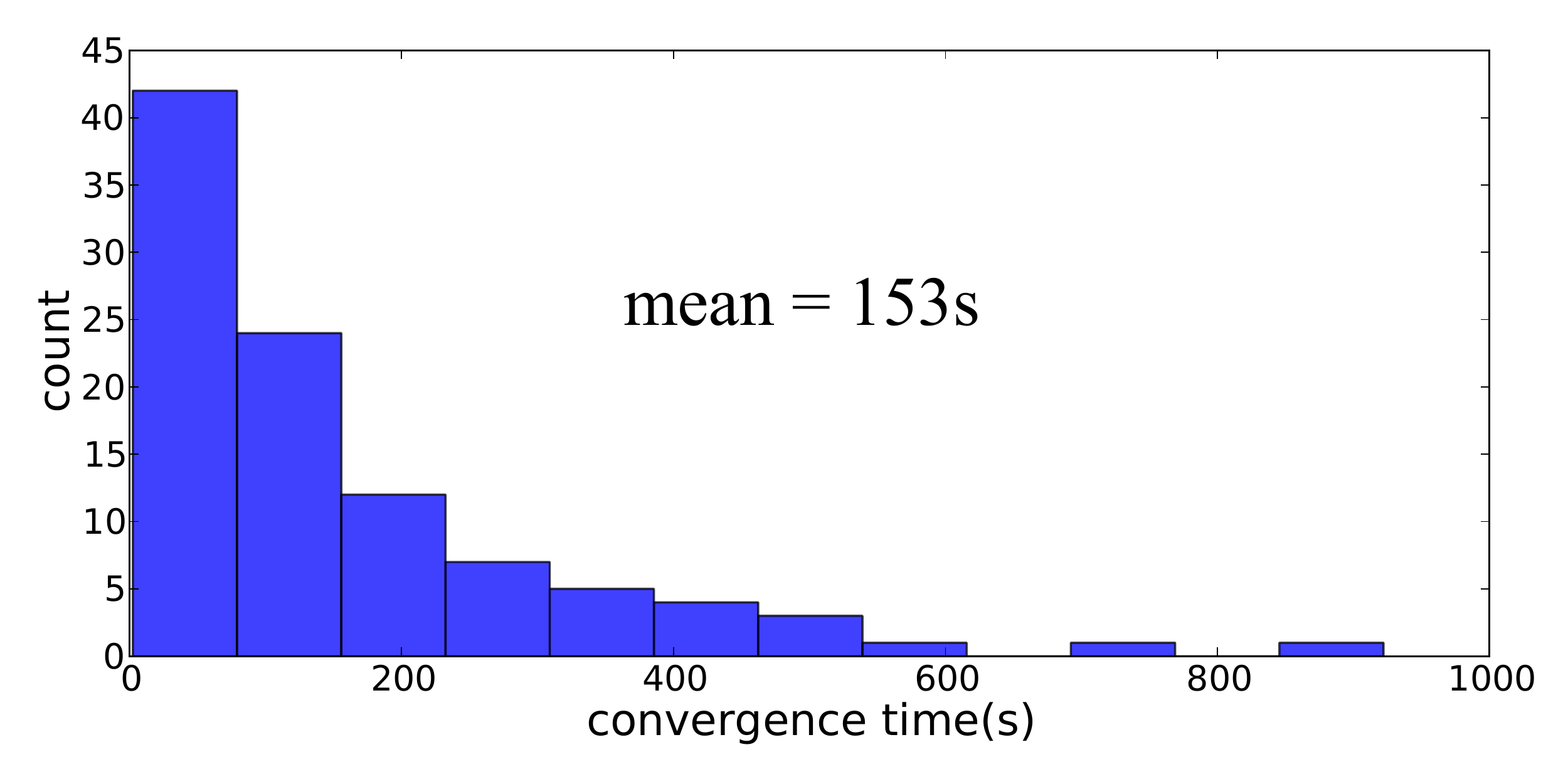} 
     \subcaption{}
     \label{fig:sud_b}
   \end{subfigure}
   \\
   \begin{subfigure}[]{0.5\textwidth}
     \includegraphics[width=\textwidth]{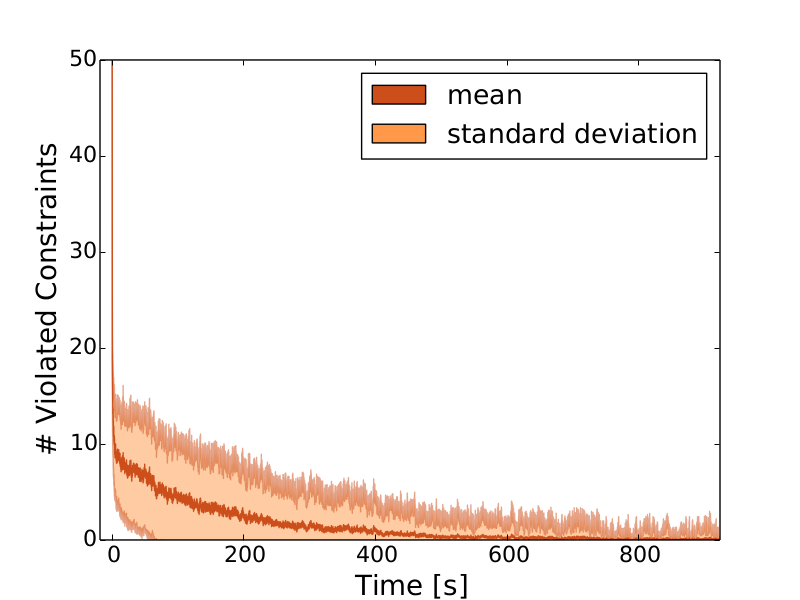} 
     \subcaption{} 
     \label{fig:sud_c}
   \end{subfigure}
   \quad
   \begin{subfigure}[]{0.5\textwidth}
     \includegraphics[width=\textwidth]{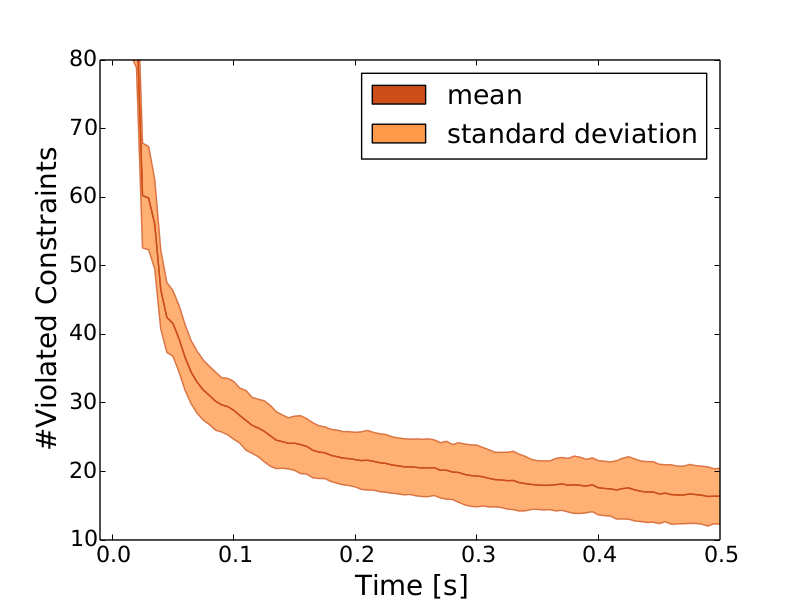} 
     \subcaption{} 
     \label{fig:sud_d}
   \end{subfigure}
   \caption{(\subref{fig:sud_a}) A 9*9 hard Sudoku instance. Each square is represented by one WTA with 9 excitatory populations, which receives oscillatory inhibition. (\subref{fig:sud_b}) Histogram of the convergence time to the unique solution in 100 trials starting from different initial conditions. (\subref{fig:sud_c}) Mean and standard deviation of the number of inequality constraints violated by the Sudoku network at each time point across the 100 trials. (\subref{fig:sud_d}) A zoom-in of the first 0.5\,s in \subref{fig:sud_c}. }
   \label{fig:sud}
\end{figure}

The proofs are given in the supplementary material. The network thus traverses the space of possible WTA configurations in an irregular, aperiodic manner until it finds a fully consistent configuration. External inputs can clamp the states of some WTAs, i.e, force particular populations to win the competition in each inhibition cycle. In this case, the rest of the network would search for configurations that satisfy the consistency conditions, given the states of the input-clamped WTAs. We illustrate this behavior using a hard 9*9 Sudoku instance (with one unique solution) that is shown in Fig.~\ref{fig:sud_a}. The network embodying this problem uses a WTA circuit with 9 excitatory populations to represent each square. The index of the winning population in each WTA encodes the number in the underlying square. The Sudoku constraints are implemented by coupling each square/WTA to every other square/WTA in the same row, column, or 3*3 block with a pair-wise connection scheme that encodes an inequality constraint: each excitatory population in one square/WTA is connected to all the excitatory populations in the other square/WTA {\it except} the excitatory population having the same index (Fig.~\ref{fig:arch_desc_d} shows an inequality constraint in the binary WTA case). Each WTA will then try to force all the other WTAs in the same row, column, and 3*3 block to encode a different number. External input forces every WTA that represents a pre-filled square to always encode the pre-filled number. For example, the top left WTA/square receives external input so that population $8$ always wins in each inhibition cycle.

Figures~\ref{fig:sud_c},~\ref{fig:sud_d} show that the network quickly yields reasonably good solutions that only violate a few constraints, and on average tends to occupy better configurations the longer it is allowed to run. The network is able to escape from locally optimal configurations and in all trials, it manages to find and stabilize at the solution to the Sudoku problem. Fig.~\ref{fig:sud_b} shows the convergence times when using oscillatory inhibition rhythms with an average frequency of $50\,Hz$.

\subsection*{Irreconcilable consistency conditions and the sampling analogy}
\begin{figure}
  \centering
   \begin{subfigure}[b]{0.5\textwidth}
     \includegraphics[width=\textwidth]{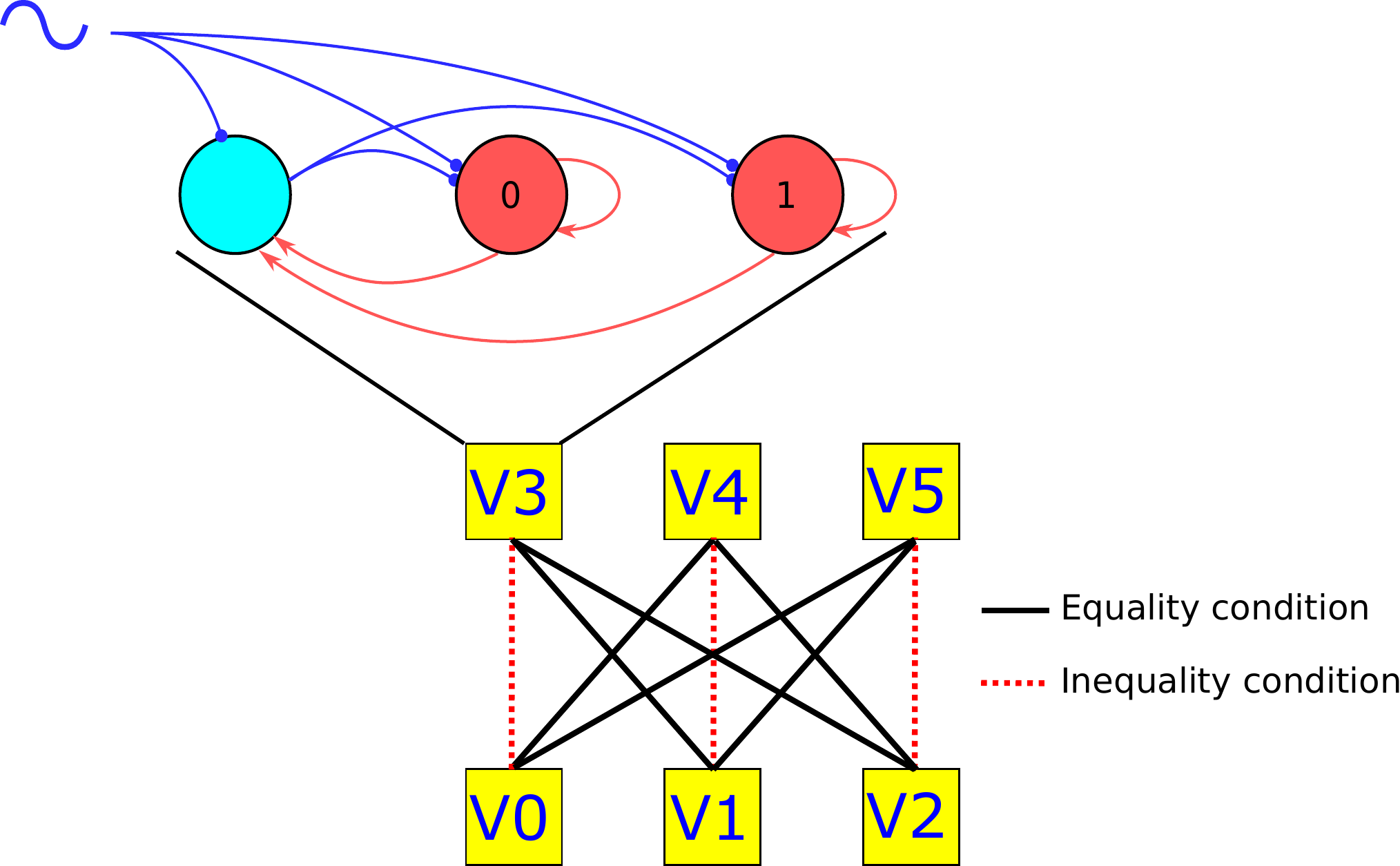} 
     \subcaption{} 
     \label{fig:sampling_a}
   \end{subfigure} \hfill
   \begin{subfigure}[b]{0.47\textwidth}
     \includegraphics[width=\textwidth]{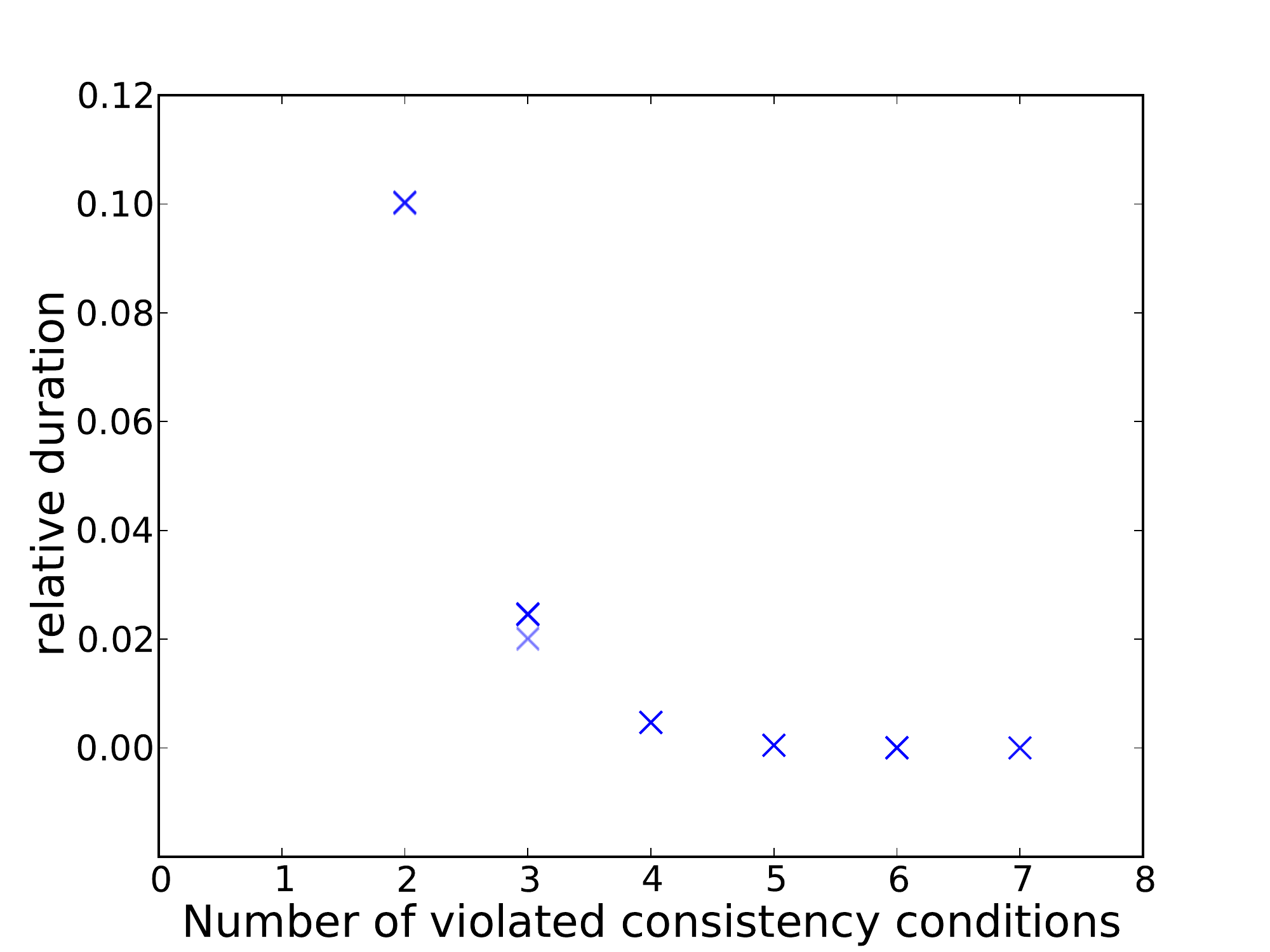} 
     \subcaption{}
     \label{fig:sampling_b}
   \end{subfigure} \\
   \begin{subfigure}[b]{0.47\textwidth}
     \includegraphics[width=\textwidth]{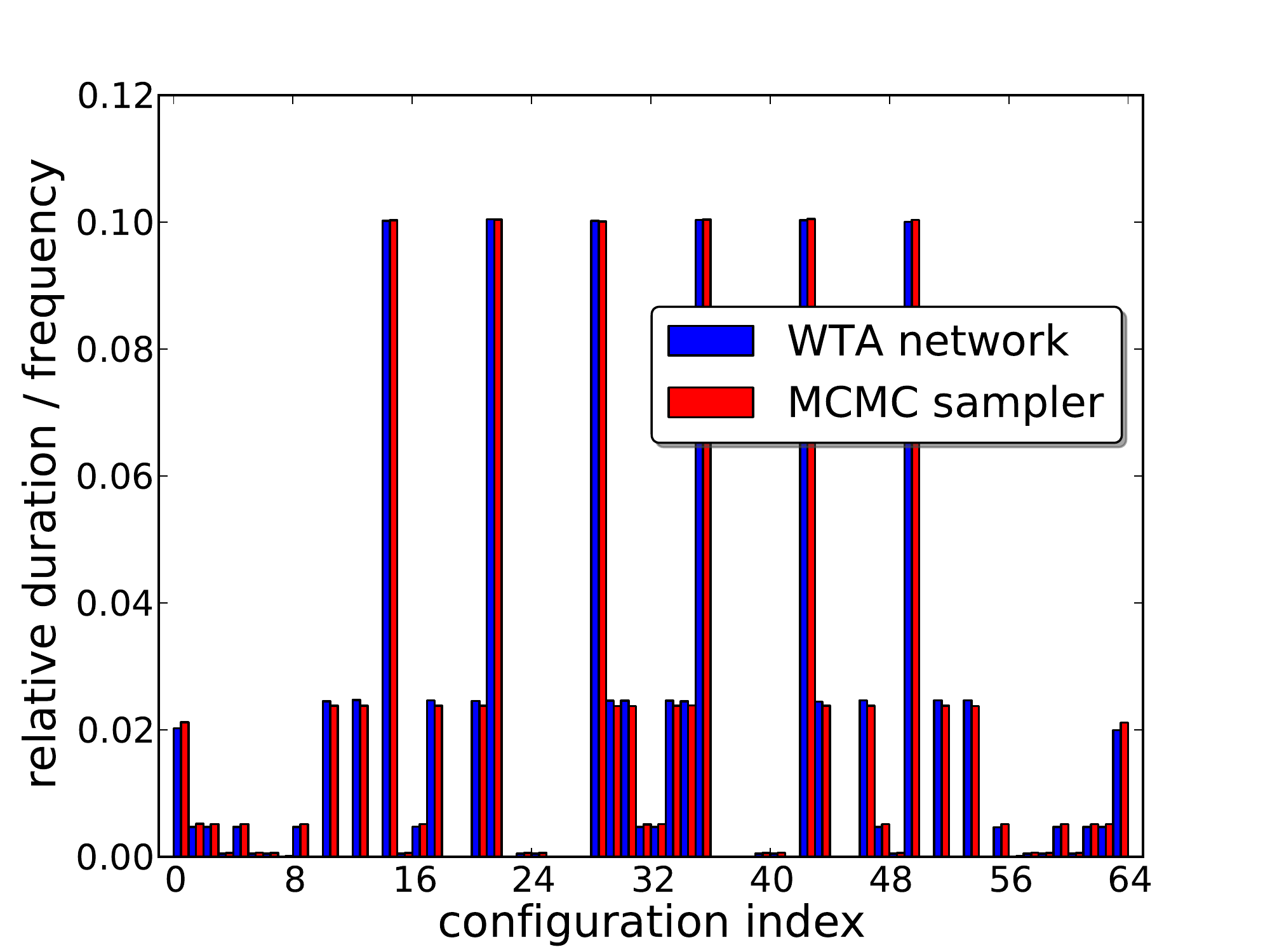} 
     \subcaption{}
     \label{fig:sampling_c}
   \end{subfigure}
   \begin{subfigure}[b]{0.52\textwidth}
     \includegraphics[width=\textwidth]{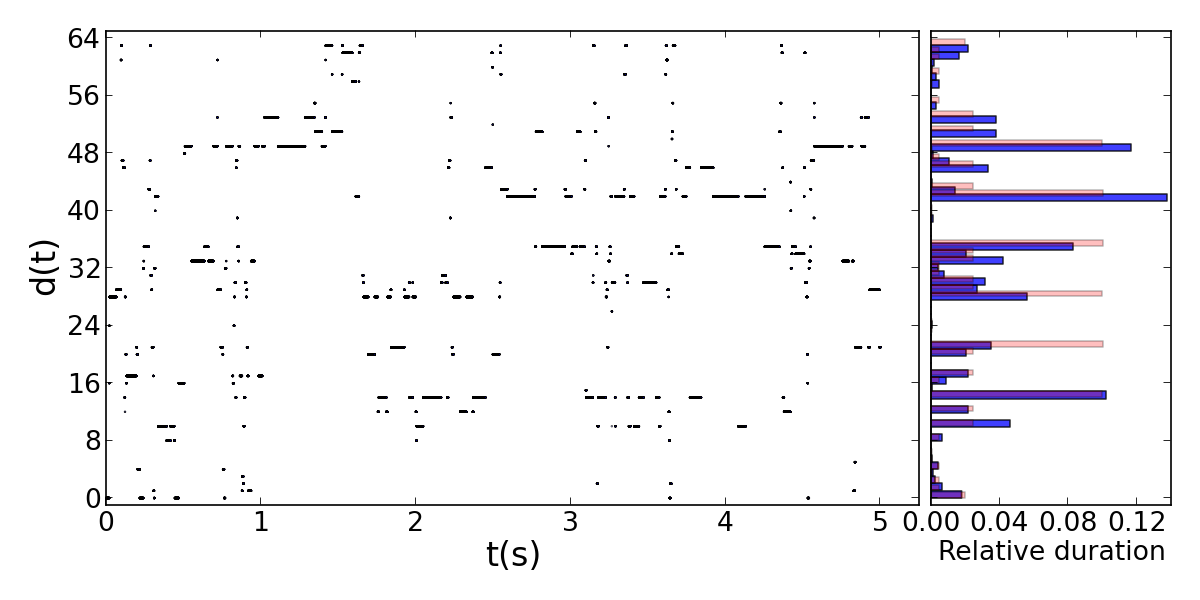} 
     \subcaption{}
     \label{fig:sampling_d}
   \end{subfigure}

\caption{(\subref{fig:sampling_a}) Topology of a network  with irreconcilable consistency conditions. Each square corresponds to a binary WTA. (\subref{fig:sampling_b}) Normalized duration of time spent by the network in each of the 64 configurations as a function of the number of violated consistency conditions. Each cross denotes one configuration. (\subref{fig:sampling_c})  Normalized durations of the 64 network configurations obtained by simulating the actual network for $1\times10^6$\,seconds and the relative frequencies of occurrence of these configurations in a sequence of $10^8$ samples generated by a stochastic MCMC operator constructed to approximate the network trajectory.  The configuration index is the decimal value of the binary string encoding the states of $V0$ to $V5$. (\subref{fig:sampling_d}) Trajectory of the dynamical variable $d(t)$ for the network in \subref{fig:sampling_a} for the first $5$ seconds (on the left), with the normalized duration of time spent in the different configurations during those $5$ seconds (on the right, blue bars). As the simulation progresses, the normalized time spent in the different configurations approaches the normalized duration results of \subref{fig:sampling_c} (reproduced on the right, in light red).}
\label{fig:sampling}
\end{figure}

According to proposition 1, if the consistency conditions encoded in the coupling connections can not all be simultaneously satisfied, the network will never settle in one configuration. This is the case for the network shown in Fig.~\ref{fig:sampling}. Figure ~\ref{fig:sampling_b} shows that the network tends to spend more time in configurations that violate fewer consistency conditions. Figure ~\ref{fig:sampling_c} shows the relative duration of time the network spends in each configuration. The plot in Fig.~\ref{fig:sampling_c} can be interpreted as the plot of a probability distribution over the possible configurations of the network. We take this stochastic interpretation further by approximating the network dynamics by a Markov chain Monte Carlo (MCMC) sampling process. Based on the network topology, we construct a stochastic Markovian transition operator $T$ that approximates the network trajectory (see Methods section).

We compare the normalized duration of time the trajectory of $d(t)$ spends in a particular configuration to the normalized frequency of occurrence of this particular configuration in the sequence of samples generated by the stochastic transition operator $T$. The results are shown in Fig.~\ref{fig:sampling_c} for the example network. The two sets of statistics are similar even though they were generated by two widely different classes of systems: one that is deterministic, continuous-valued and running in continuous time, and the other stochastic, discrete-valued and running in discrete time steps. The sampling analogy is reinforced by the aperiodicity of $d(t)$ which yields an irregular trajectory of the network state that is akin to a truly stochastic trajectory as shown in Fig.~\ref{fig:sampling_d}. The difference between two probability distribution $p$ and $q$ can be quantified using Kullback-Leibler(KL) divergence:
$$KL(p,q) = \sum\limits_{x_i}p(x = x_i)log\left(\frac{p(x=x_i)}{q(x=x_i)}\right)$$
For the network shown in Fig.~\ref{fig:sampling}, $KL(p_{network},p_{MCMC}) = 9.0*10^{-4} nat$. In the supplementary material, we show that the stochastic approximation is still good for various network sizes and inter-WTA connection densities (see Supplementary Fig.~S1) and analyze the properties of the MCMC operator as well as discuss in more detail the assumptions underlying the stochastic approximation. 

Of course, given the statistics of the network trajectory, it is trivial to construct an MCMC operator that has these statistics as a stationary distribution using any of the standard sampling methods such as Gibbs sampling. The MCMC operator described in this section, however, was constructed a-priori, i.e, before simulating the network, as an approximation of the network trajectory.

\subsubsection*{Effect of noise and non-zero coherence}
\begin{figure}
   \begin{subfigure}[b]{0.5\textwidth}
     \includegraphics[width=\textwidth]{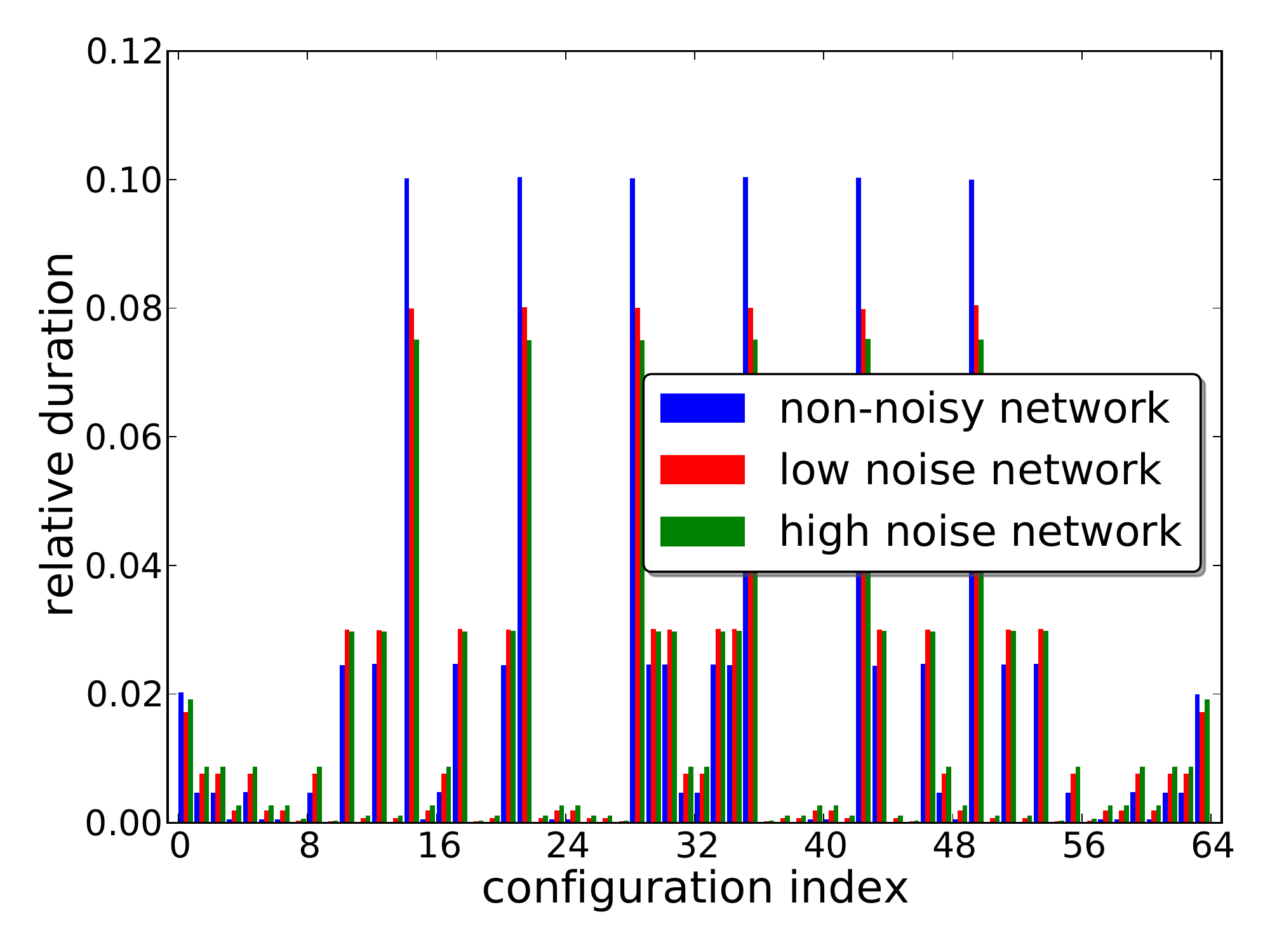} 
     \subcaption{} 
     \label{fig:coh_a}
   \end{subfigure}
   \hfill
   \begin{subfigure}[b]{0.5\textwidth}
     \includegraphics[width=\textwidth]{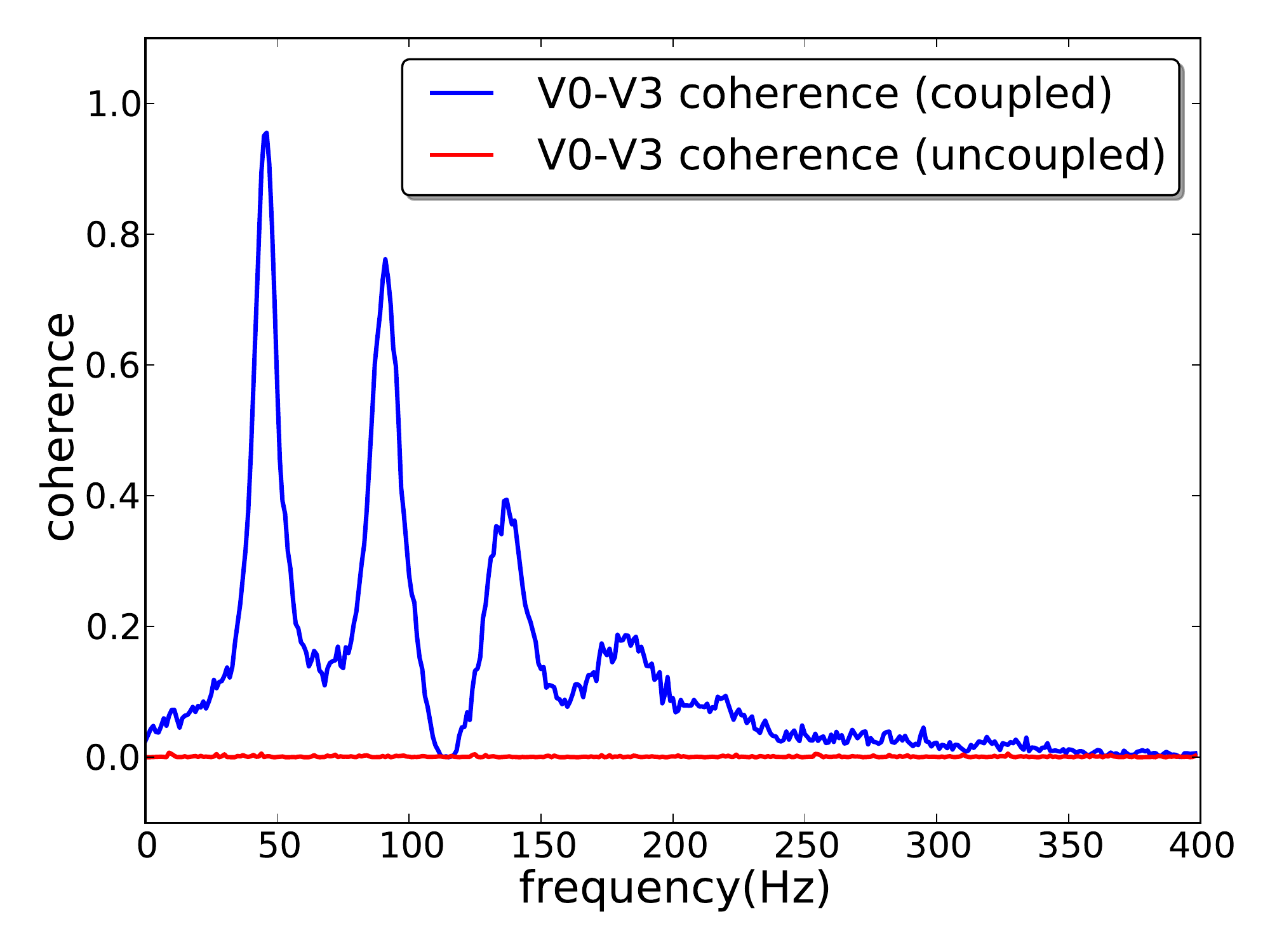} 
     \subcaption{}
     \label{fig:coh_b}
   \end{subfigure}
   \\
   \begin{subfigure}[l]{0.5\textwidth}
     \includegraphics[width=\textwidth]{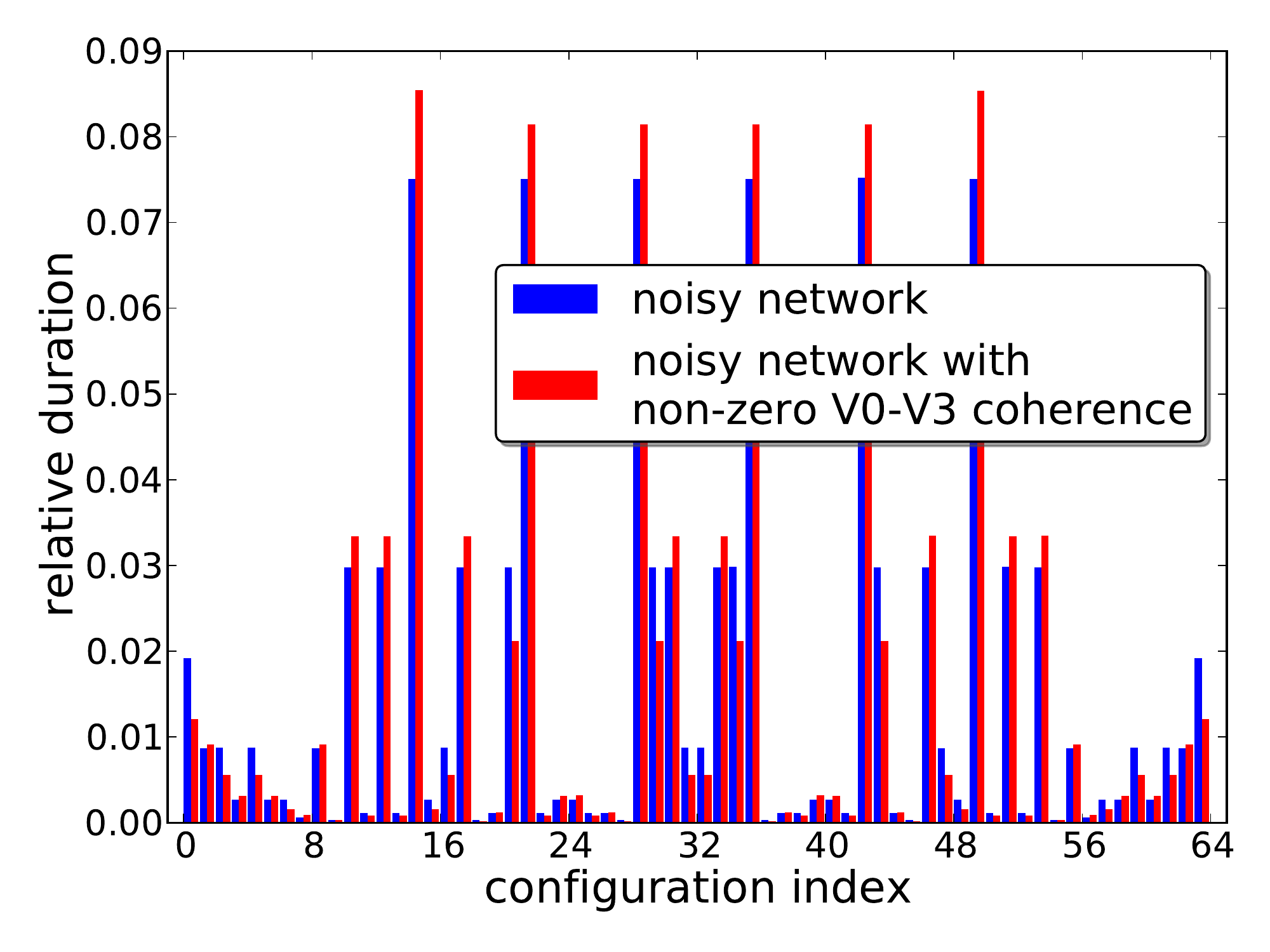} 
     \subcaption{} 
     \label{fig:coh_c}
   \end{subfigure}
   \hfill
   \begin{subfigure}[l]{0.5\textwidth}
     \includegraphics[width=\textwidth]{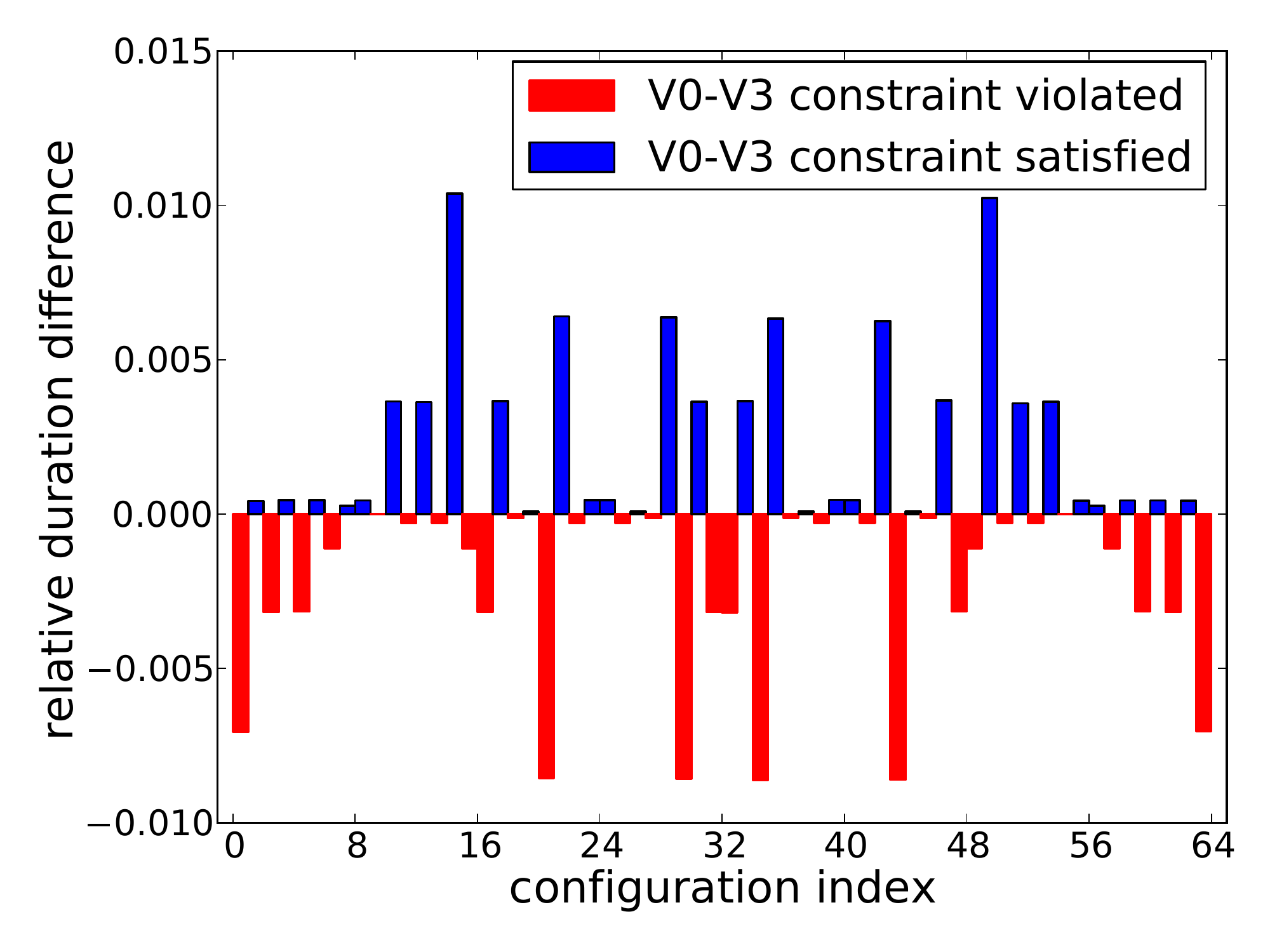} 
     \subcaption{} 
     \label{fig:coh_d}
   \end{subfigure}

   \caption{(\subref{fig:coh_a}) Statistics of the trajectory of the network in Fig.~\ref{fig:sampling_a} in the non-noisy, low noise, and high noise cases. (\subref{fig:coh_b}) Mean square coherence between the oscillatory inhibition in variables V0 and V3 in the high noise case when their phases are coupled and when they are uncoupled. (\subref{fig:sud_c}) statistics in the high noise case when V0 and V3 are phase coupled and when they are uncoupled. (\subref{fig:coh_d}) Change in the relative (normalized) duration of each network configuration when V0-V3 become phase coupled. Data was obtained from \subref{fig:coh_c}. Blue bars indicate configurations in which the V0-V3 consistency condition is satisfied while red bars indicate configurations in which the consistency condition is violated.}
   \label{fig:coh}
\end{figure}

The deterministic networks we present exploit the continuously shifting phase relations between different WTAs in order to explore the space of possible network configurations. We added white Gaussian noise to the input of all LTUs and random fluctuations to the phases of all inhibitory rhythms (see Methods section). Figure ~\ref{fig:coh_a} shows that noise `smears out' the distribution of relative durations by biasing the trajectory towards low-duration states and away from high-duration states. This smearing out effect is more pronounced for larger random fluctuations. This is reminiscent of how higher noise/temperature in a thermodynamical system acts to increase the system entropy by making the probability distribution over the system states more uniform. There is a crucial difference between these networks and thermodynamical systems, however: At the zero noise (zero temperature) level, a thermodynamical system quickly freezes at a single state while our networks will continue to explore the configuration space. 

Gamma rhythms in distant neural assemblies often exhibit non-zero but still imperfect (non-unity) coherence~\cite{Bosman_etal12}. We model this phenomenon by coupling the phases of the inhibitory rhythms in different WTAs according to the Kuramoto model~\cite{Strogatz00} in order to realize phase relations that are consistently near zero. This phase coupling was introduced between variables/WTAs V0 and V3 in the network in Fig.~\ref{fig:sampling_a} and all network components were perturbed by random fluctuations. Figure~\ref{fig:coh_d} shows that V0-V3 coherence increases the duration of all configurations in which the V0-V3 consistency condition is satisfied and decreases the duration of all configurations in which it is not. Increased coherence between two WTAs thus leads to stronger interaction and makes it more difficult to violate the consistency conditions between them.

\subsection*{Learning the consistency conditions}
Like any connectionist architecture, the knowledge in the networks we describe is encoded in the inter-WTA connections. These connections represent the consistency conditions which, through the network dynamics, induce a ``probability distribution'' over the network configurations. If made plastic, the strengths of these connections will be continuously modulated by the changing network state. By forcing a particular configuration on the network, external input can thus reorganize the pattern of these inter-WTA connections so as to maximize the consistency of the input-imposed configurations as interpreted according to the pattern of inter-WTA connections. A central question in such a learning scheme is how the plasticity rule can distinguish between configurations that are input-imposed, and thus should be learned, and configurations that arise naturally when the network is running freely. This central question also arises in stochastic, sampling-based connectionist architectures that learn probability distributions by example such as Restricted Boltzmann Machines~\cite{Ackley_etal85}. 

To distinguish an input-imposed configuration as a configuration that should be learned, i.e, that should serve as a model for consistent configurations, we have external input synchronize the activity of the WTA circuits it targets so that the oscillatory inhibition in these WTA circuits has a common frequency and a zero phase difference. Synchronization is thus a dynamical marker for configurations that should be learned. We use a bistable Hebbian plasticity rule that acts on the weights of the inter-WTA coupling connections. A drift term pushes the synaptic weight to one of two stable values: high or low (see Methods section). 

In a rate based network, we are unable to capture the way gamma-band synchronization enhances plasticity by forcing neurons in multiple WTA circuits to spike within narrow time windows~\cite{Axmacher_etal06}. However, the plasticity rule we use results in the same functional effect. In order for a depressed connection to potentiate, both pre- and postsynaptic rates need to be large at the same time for many consecutive inhibition cycles in order to overcome the bistability drift. This will only reliably happen if the oscillatory inhibition in the pre- and post-synaptic WTA circuits have a measure of synchrony where their phase difference is around zero for many cycles so that the peaks of excitatory activity in the WTA circuits coincide. 

\subsection*{Learning in a model of perceptual multi-stability}
In this section, we describe a network that is able to learn a consistency model by example and then reproduces a perceptual multi-stability phenomenon when faced with ambiguous inputs. The particular form of perceptual multi-stability we consider is binocular rivalry which has been well studied experimentally \cite{Mamassian_Goutcher05} as well as in theory \cite{Gershman_etal12}.
In binocular rivalry experiments, each eye is shown a differently oriented grating. The subject's perception then continuously switches between the two orientations~\cite{Mamassian_Goutcher05}. 

\paragraph{Network Model} The network is composed of $n_{hid} =6$ hidden and $n_{in} = 6$ input WTA circuits that each undergo oscillatory inhibition.  Each WTA has six competing excitatory populations,  $n_c =6$, and can  take one of six different states (orientations). $n_c =6$ implies that each excitatory population codes for a range of orientations of $\frac{180^\circ}{6} = 30^\circ$. 

\begin{figure}
  \centering
     \includegraphics[width=\textwidth]{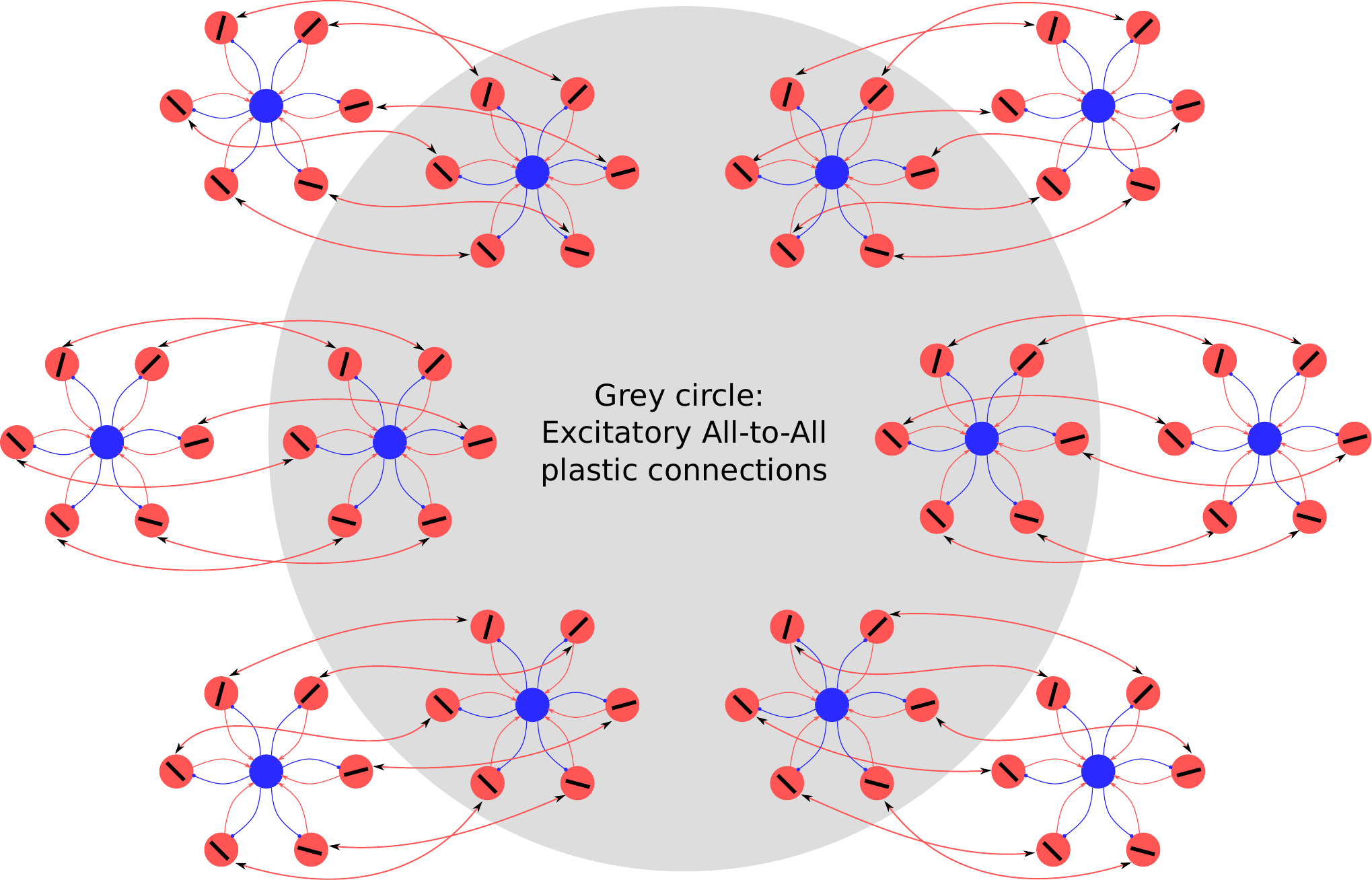} 
     \caption{Network architecture for the perceptual multi-stability task. Hidden WTAs are inside the grey circle while input WTAs are outside. All explicitly drawn connections are fixed connections. The six hidden WTA circuits have all-to-all plastic connectivity, i.e, each excitatory population in one hidden WTA connects to each excitatory population in the five other hidden WTA circuits.  Local oscillatory inhibition in each WTA and the recurrent excitatory connections are not shown.}
\label{fig:multinet}
\end{figure}	

The full network connectivity is illustrated in Fig.~\ref{fig:multinet}. Input and hidden WTA circuits are set up in pairs.
Each input WTA connects to one hidden WTA so that populations of similar orientations are bidirectionally coupled. The connectivity between hidden WTA circuits is all-to-all and initialized using random weights: $w^i_{init} \in [w_{min},w_{max}]$ that are \emph{plastic} and follow a bistable Hebbian plasticity rule(see Methods section for more details).

\paragraph{Training and Testing} The simulations we run have two distinct phases, a training phase in which we cycle through consistent inputs by clamping the input WTAs to the same orientation, and a testing phase in which we clamp a subset (or all) of the input WTAs at various orientation patterns and decode the activity of the hidden WTA circuits.  During the two phases we use \emph{the same synaptic rule and parameters}; learning is effectively enabled/disabled by providing/withholding a synchronizing oscillatory input to the inhibitory oscillators of all WTAs (see Methods for further details). 

\begin{figure}
  \centering
  \label{fig:bistab}
     \begin{subfigure}[]{0.33\textwidth}
     \includegraphics[width=\textwidth]{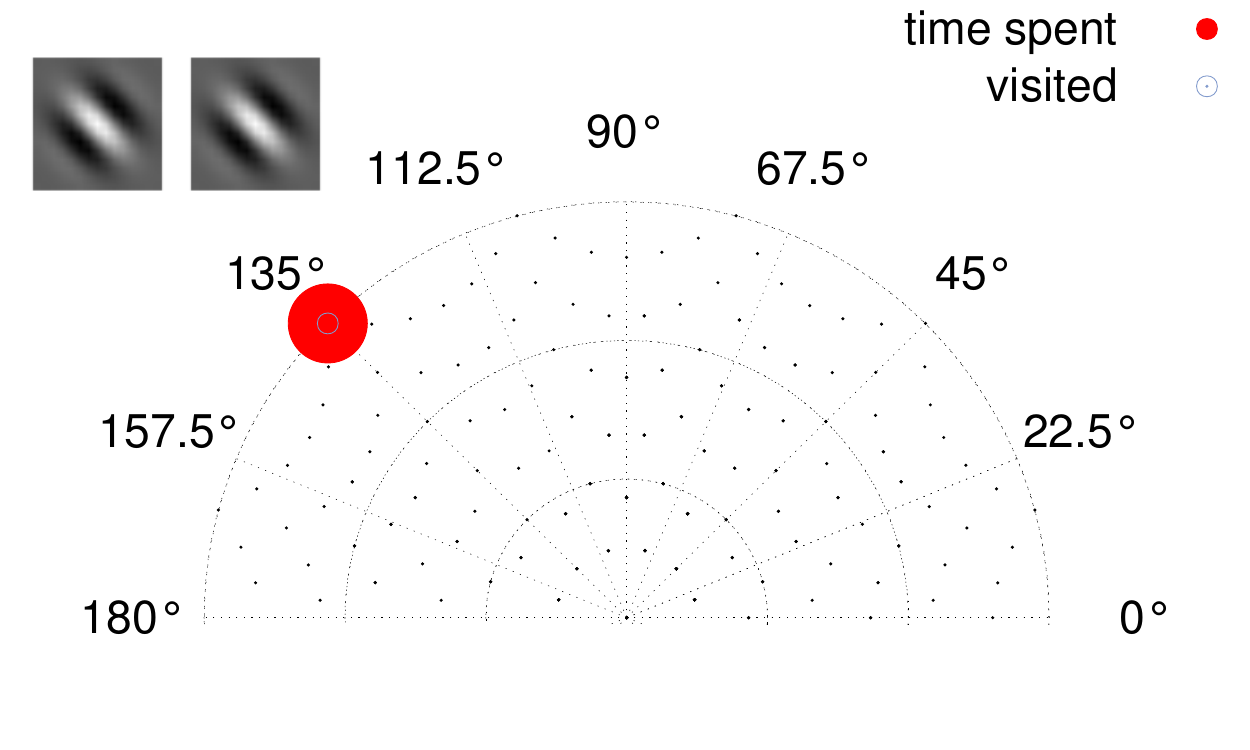} 
     \subcaption{Trained network, unambiguous input}
     \label{fig:bistab_st}
   \end{subfigure}
   \quad
   \begin{subfigure}[]{0.33\textwidth}
     \includegraphics[width=\textwidth]{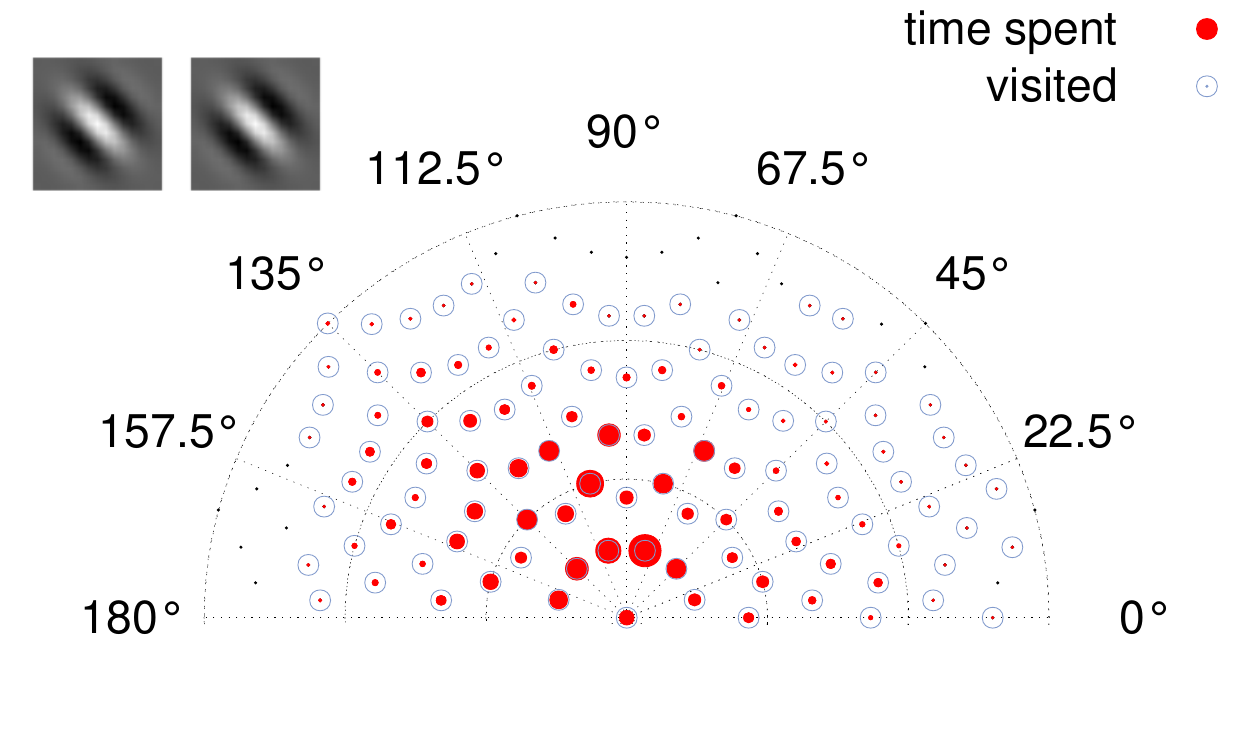} 
     \subcaption{Untrained network, unambiguous input}
     \label{fig:bistab_sr}
   \end{subfigure}\\
   \begin{subfigure}[]{0.33\textwidth}
     \includegraphics[width=\textwidth]{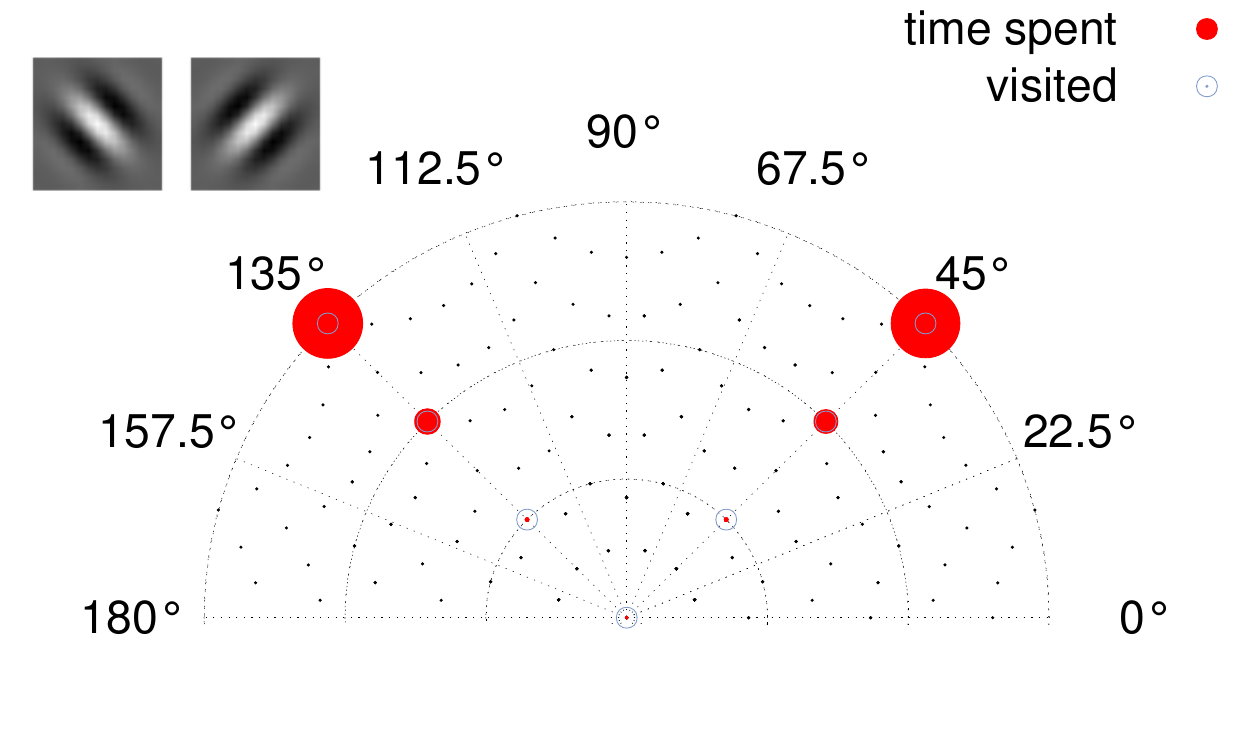} 
     \subcaption{Trained network, 2 direction ambiguous input} 
     \label{fig:bistab_2t}
   \end{subfigure}
   \quad
   \begin{subfigure}[]{0.33\textwidth}
     \includegraphics[width=\textwidth]{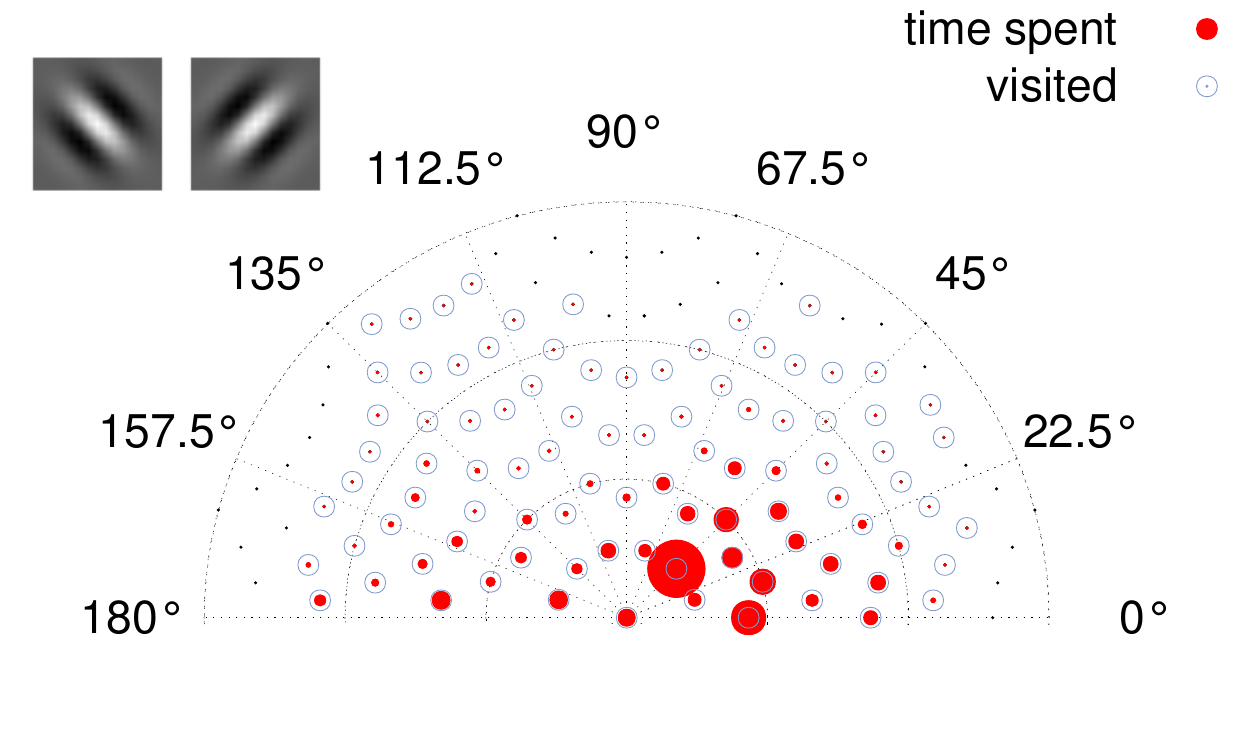} 
     \subcaption{Untrained network, 2 direction ambiguous input}
     \label{fig:bistab_2r}
   \end{subfigure}\\
   \begin{subfigure}[]{0.33\textwidth}
     \includegraphics[width=\textwidth]{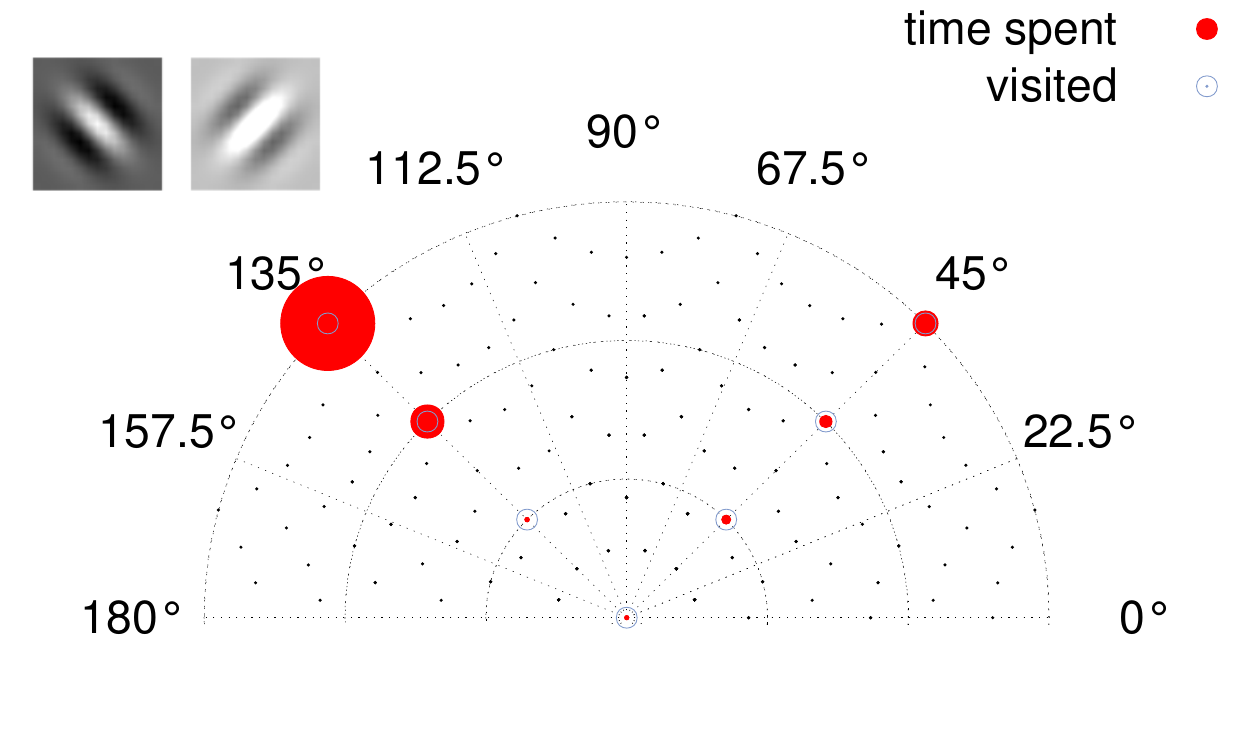} 
     \subcaption{Trained network, biased ambiguous input}
     \label{fig:bistab_bt}
   \end{subfigure}
   \quad
   \begin{subfigure}[]{0.33\textwidth}
     \includegraphics[width=\textwidth]{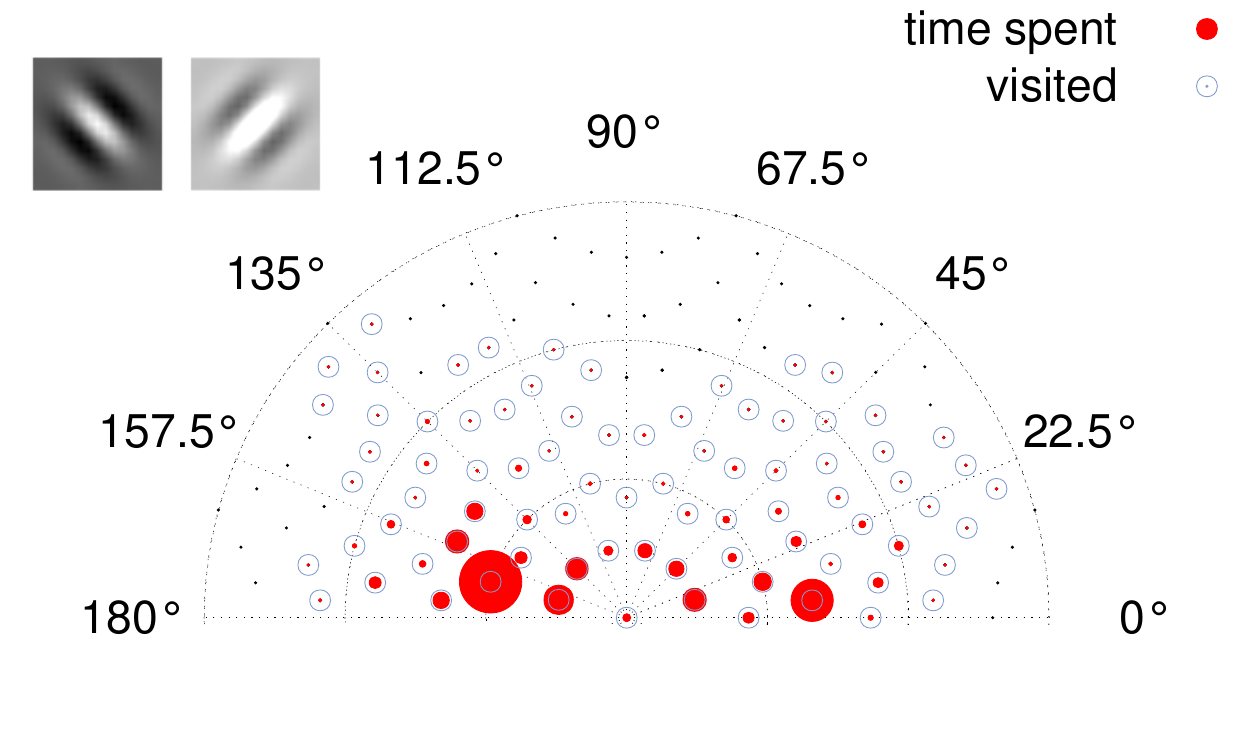} 
     \subcaption{Untrained network, biased ambiguous input}
     \label{fig:bistab_br}
   \end{subfigure}\\
      \begin{subfigure}[]{0.33\textwidth}
     \includegraphics[width=\textwidth]{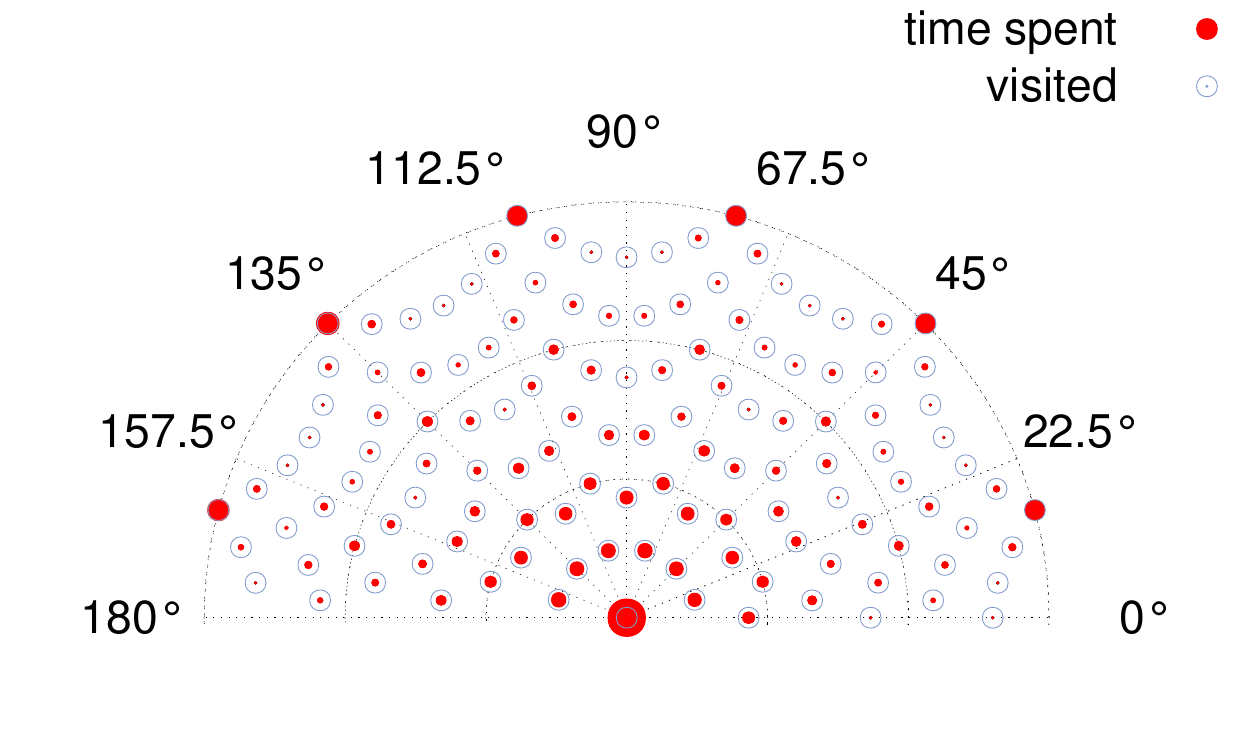} 
     \subcaption{Trained network, 6 direction ambiguous input}
     \label{fig:bistab_6t}
   \end{subfigure}
   \quad
   \begin{subfigure}[]{0.33\textwidth}
     \includegraphics[width=\textwidth]{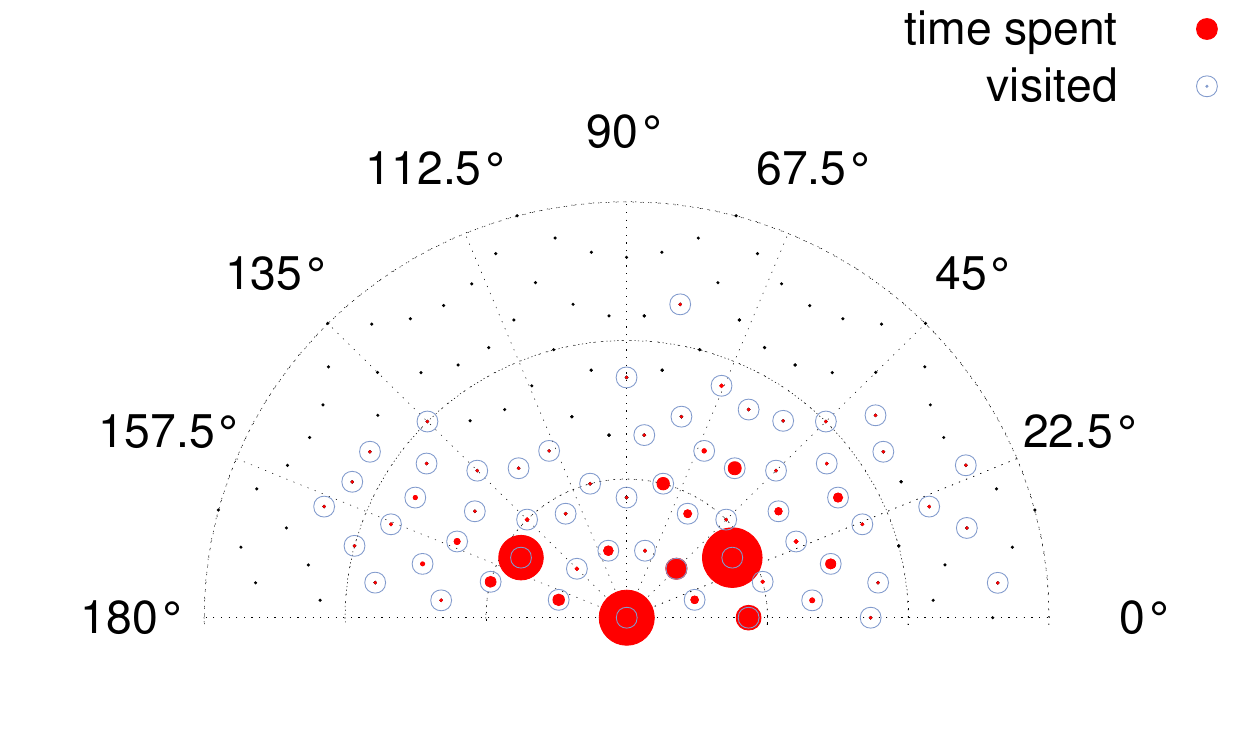} 
     \subcaption{Untrained network, 6 direction ambiguous input}
     \label{fig:bistab_6r}
   \end{subfigure}\\
   \caption{The ``probability distribution'' of the orientation vector decoded from the hidden WTAs in the network in Fig.~\ref{fig:multinet} under different conditions.
   Black dots indicate states that can be encoded by the network, and blue empty circles indicate visited states.
    The area of the filled red circle at each visited location is proportional to the time spent at that state (scale differs between plots). 
    The inset Gabor patches indicate the stimuli presented to the left and right eye in analogous experimental settings.
   (\subref{fig:bistab_st}) A network trained on consistent inputs propagates the clamped angle of one input WTA to the other WTA circuits and stably represents that angle.
   (\subref{fig:bistab_sr}, \subref{fig:bistab_2r}, \subref{fig:bistab_br}, \subref{fig:bistab_6r}) An untrained network spends most of the time in ``low confidence states'' where the hidden WTA circuits encode different angles in all input cases.
   (\subref{fig:bistab_2t}) The trained network switches between two interpretations of an ambiguous input. 
   (\subref{fig:bistab_bt}) Increasing the number of input WTA circuits encoding $135^\circ$ to two makes that interpretation more probable compared to the $45^\circ$ interpretation
   (\subref{fig:bistab_6t}) The trained network under fully conflicting input preferentially visits consistent states, where all populations represent the same direction. ``Low confidence'' states are also visited. 
 }
\label{fig:bistab}
\end{figure}

To visualize the network behavior, we sum the preferred orientation vectors of the winning excitatory populations in the hidden WTAs to obtain a vector whose angle represents the current guess of the network at the true orientation which it `senses' through the input WTA circuits. The vector's magnitude represents the confidence of the network in this orientation estimate (see Methods section for more details). 
Fig.~\ref{fig:bistab} illustrates the trained and untrained network responses in various input cases: unambiguous input in Figs.~\ref{fig:bistab_st} and ~\ref{fig:bistab_sr} (one input WTA clamped at $45^\circ$ orientation), ambiguous input in Figs.~\ref{fig:bistab_2t},\ref{fig:bistab_2r},\ref{fig:bistab_bt}, and~\ref{fig:bistab_br} (one input WTA clamped at $45^\circ$ and one or two other input WTAs clamped at $135^\circ$), and a fully conflicting input in Figs.~\ref{fig:bistab_6t} and \ref{fig:bistab_6r} (each input WTA is clamped at a different angle). 

\begin{figure}
\centering
   \begin{subfigure}[]{0.33\textwidth}
     \includegraphics[width=\textwidth]{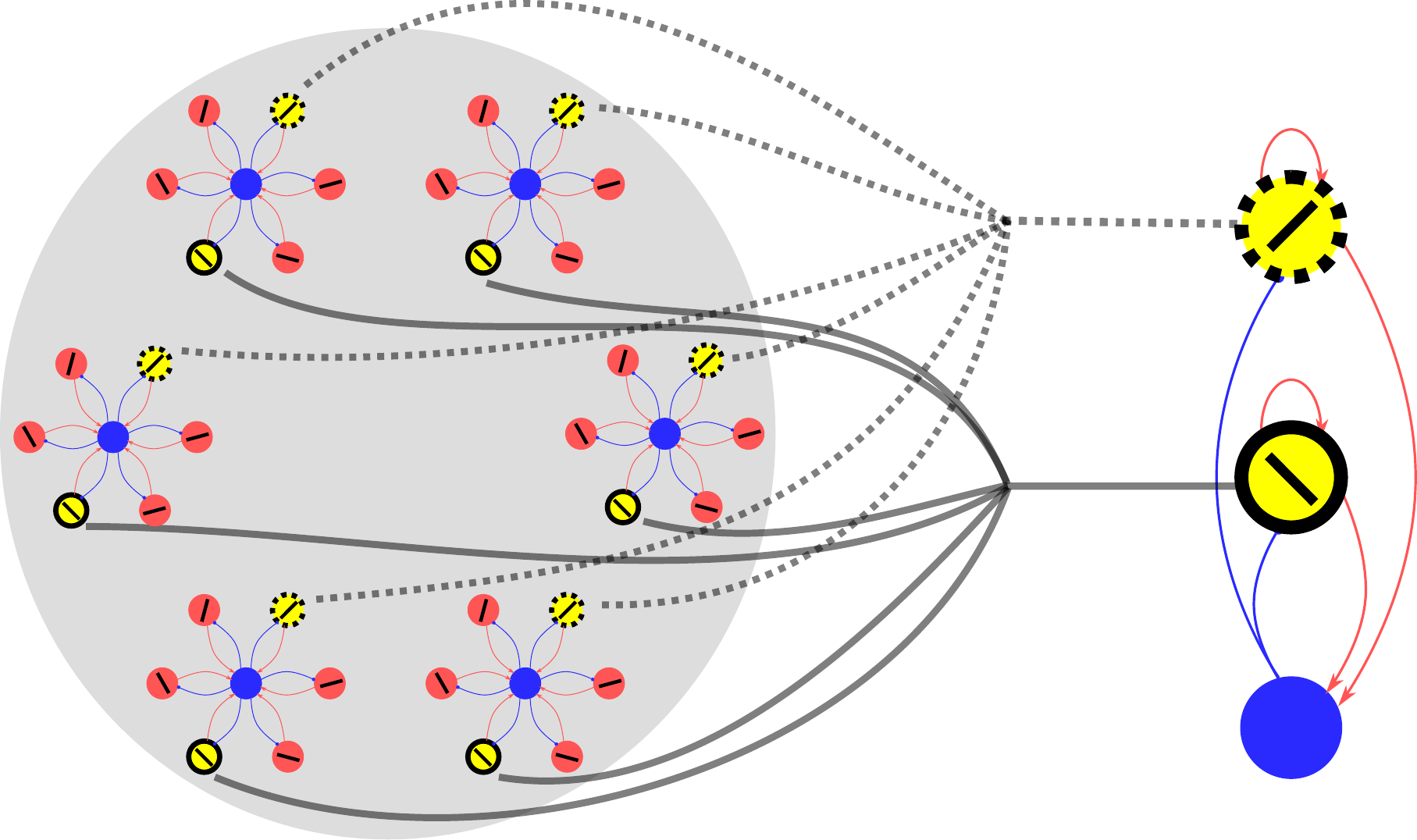} 
     \subcaption{} 
     \label{fig:readout}
   \end{subfigure}
   \quad
   \quad
   \begin{subfigure}[]{0.4\textwidth}
     \includegraphics[width=\textwidth]{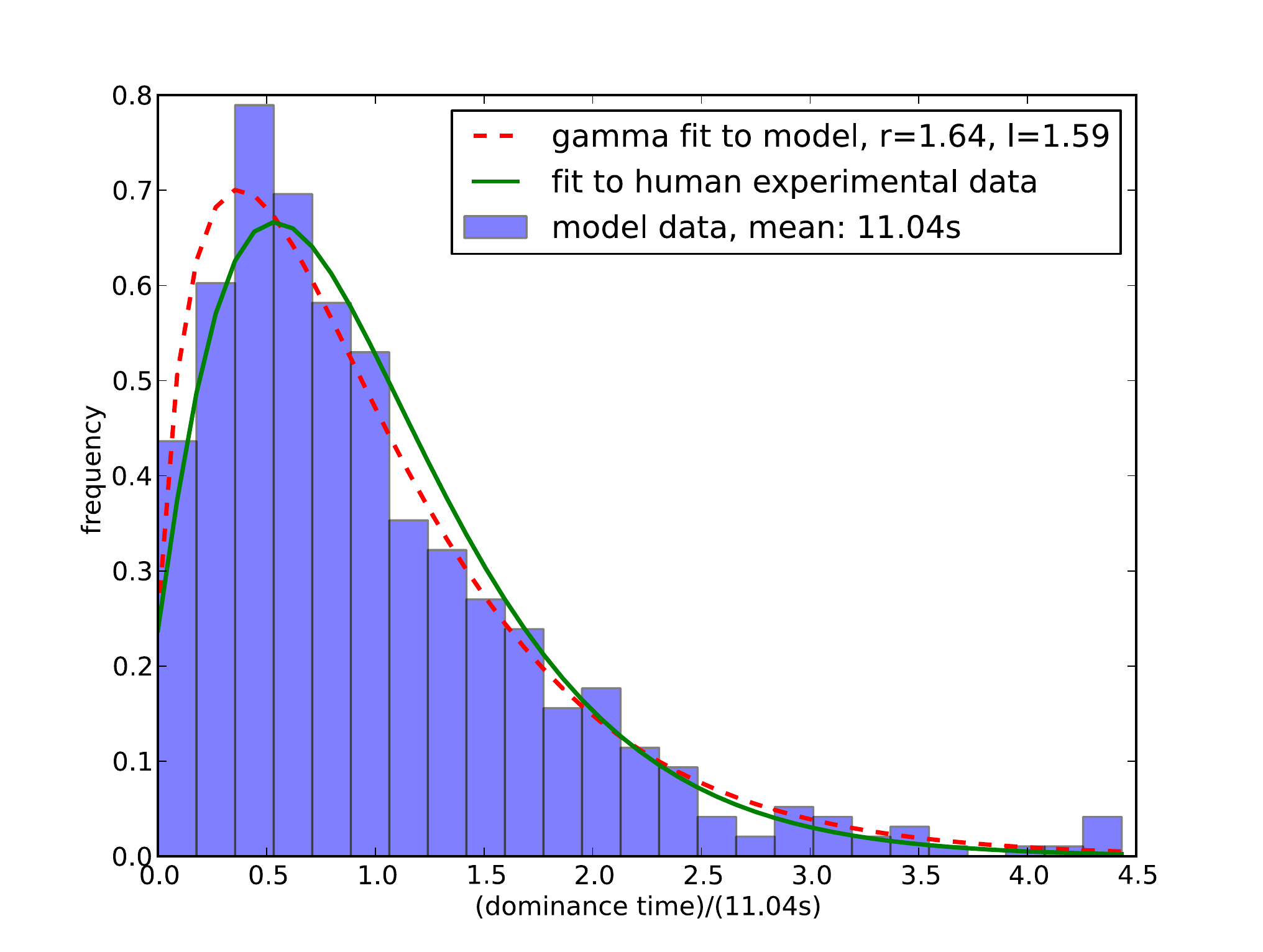}
     \subcaption{} 
     \label{fig:dt}
   \end{subfigure}
   \caption{(\subref{fig:readout})Readout WTA performing the perceptual decision similar to \cite{Wong_etal07}. (\subref{fig:dt})The perceptual switching times produced by our model fit a gamma distribution that largely agrees with experimental data from \cite{Mamassian_Goutcher05}.}
   \label{fig:switchStats}
\end{figure}

To be able to make a quantitative comparison to human experiments, we investigated the perceptual switching times our model produces when presented with an ambiguous stimulus. We added an `integrative' component to the system  based on \cite{Wong_etal07} in which a bistable decision task for random dot stimuli is modelled: A bistable readout attractor network receives time-varying inputs and settles into one attractor state. The time varying input is produced by the hidden WTAs and is slowly integrated by the readout attractor network. The bistable attractor's population with the higher activity corresponds to the current percept, see Fig.~\ref{fig:switchStats}.

\section*{Discussion}
\label{sec:discussion}

Several theories postulate that the key computational machinery of sensory and association cortices is devoted to solving best-match problems~\cite{Rumelhart_McClelland86} where a number of interacting areas try to agree on a global interpretation that best explains sensory data while being consistent with an internal model, or a prior, of the environment~\cite{Berkes_etal11,Friston03}. We showed that oscillatory inhibition that gives rise to rhythmic modulation of firing rate and sensitivity to external inputs in the local circuitry allows a neural network to solve best-match problems. If no fully consistent network configuration exists, the network trajectory wanders aperiodically and can be approximated by a stochastic sampling process. We have shown that deterministic networks can reproduce effects that were commonly modeled using stochastic networks such as perceptual multi-stability and the explaining away effect (see supplementary Fig.~S2). Our results show that physical noise mechanisms are not strictly necessary for networks to explore complex configuration spaces without being trapped by locally optimal configurations.

Increased Gamma-band coherence between distant neural groups has been found to increase the strength of their mutual influence~\cite{Bosman_etal12,Fries05} making the coherence level of the gamma oscillations a candidate mechanism for quickly modulating the effective strength of anatomical connections (the ``communication through coherence'' hypothesis). In agreement with this effect, we have shown that if oscillatory inhibition in the WTA circuits were forced to exhibit some coherence which leads to consistent and favorable phase relations (phase differences that are around zero), communication between the coherent WTA circuits would be more effective as the consistency conditions inside the coherent set would be more difficult to violate(see Fig.~\ref{fig:coh_d}). 

We have used synchronization as a way to trigger learning. There is evidence that long-range synchronization is required for effective learning and memory formation~\cite{Weiss_Rappelsberger00,Fell_Axmacher11} and that it is a candidate mechanism for establishing associations between disparate neural groups~\cite{Miltner_etal99}. Gamma band synchronization in particular is well-suited for establishing associations through STDP mechanisms~\cite{Axmacher_etal06} as neurons fire preferentially at a particular phase of the gamma cycle (when the oscillatory inhibition is low), thereby aligning their spikes to a precision of a few milliseconds which is a suitable window for inducing synaptic potentiation or depression~\cite{Markram_etal97}. 

Our perceptual multistability model is, to our knowledge, the first model that reproduces perceptual multistability phenomena without making use of explicit stochastic dynamics. An experiment that could provide evidence as to whether our model does underlie multi-stable perception would investigate the relationship between perceptual switching times and how fast Gamma-band phase differences between fields measured at different points on the visual cortex change: faster changes in phase relations translates to a more exploratory behavior in our model as the effective network connectivity changes more rapidly and should translate to faster switching times between the different percepts.

\FloatBarrier
\section*{Methods}
\subsection*{Network equations}
The basic building block of all networks is the linear threshold unit (LTU). LTUs together with an oscillatory inhibitory population make up an oscillatory WTA as shown in Figs.~\ref{fig:arch_desc_a} and~\ref{fig:arch_desc_b}. The excitatory populations in different WTA circuits are coupled together to implement consistency conditions as shown in Fig.~\ref{fig:arch_desc_d}. The full description of a network having $M$ WTA circuits, with $N_k$ excitatory population in WTA $k$ is:
\begin{subequations}
\begin{multline}
 \tau^E \frac{d}{dt}{x}_{ij}^E(t) + x_{ij}^E(t) = f( w_{rec}^{AMPA}x_{ij}^E(t) + w_{rec}^{NMDA}s_{ij}^E(t) - w_{IE}^{GABA}x_i^I - w_{osc}^E H_i(t)  + \sigma^{LTU}\eta^E_{ij}(t) \\ + \sum\limits_{\substack{k=1 \\ k\neq i}}^{M}\sum\limits_{p=1}^{N_k}(w_{k,p\rightarrow i,j}^{AMPA}x_{kp}^E(t) + w_{k,p\rightarrow i,j}^{NMDA}(t)s_{kp}^E(t))  - T^E ) \label{eq:modela}
\end{multline}
\begin{equation}
 \tau^I \frac{d}{dt}{x}_{i}^I(t) + x_{i}^I(t) = f( - w_{osc}^I H_i(t) + \sigma^{LTU}\eta^I_i(t) +  \sum\limits_{j=1}^{N_i}(w_{EI}^{AMPA}x_{ij}^E(t) + w_{EI}^{NMDA}s_{ij}^E(t)) - T^I) \label{eq:modela}
\end{equation}
\begin{equation}
 \tau^{NMDA} \frac{d}{dt}{s}_{ij}^E(t) + s_{ij}^E(t) = x_{ij}^E(t) \label{eq:modelb}
\end{equation}

\begin{equation}
\frac{d}{dt}\phi_i = \Omega_i + \sigma^{OSC}\kappa_i\delta(\phi_i) + \sum\limits_{\substack{k=1 \\ k\neq i}}^{M}Z_{k\rightarrow i}sin(\phi_k(t)-\phi_i(t))
\label{eq:phase}
\end{equation}

\begin{equation}
 H_i(t) =   \begin{cases}
   A & \text{if \quad $\phi_i < d \cdot 2\pi$}  \\
   0       & \text{otherwise} 
  \end{cases}
\end{equation}
\end{subequations}
$x_{ij}^E(t)$ and $x_{i}^I(t)$ are the the activities of the $j^{th}$ excitatory population, and the inhibitory population in WTA $i$ respectively. $f(x)$ is the linear threshold function: $f(x)=Max(0,x)$. $\phi_i$ is the phase of the oscillatory inhibition in WTA $i$. $\phi_i$ automatically wraps around to stay in the range $[0,2\pi)$. $H_i(t)$ is the actual oscillatory inhibition waveform in WTA $i$ which, within each period, is high (with amplitude $A$) for a fraction $d < 1$ of the oscillation cycle. $x_{ij}^E(t)$ is low-pass filtered with time constant $\tau^{NMDA}$ to yield $s_{ij}^E(t)$ which is used to model slowly decaying currents mediated by $NMDA$ receptors. $\tau^E$, $T^E$ and $\tau^I$, $T^I$ are the time constants and thresholds of the excitatory and inhibitory populations respectively. $w_{k,p\rightarrow i,j}^{AMPA}(t)$ and $w_{k,p\rightarrow i,j}^{NMDA}(t)$ are the inter-WTA $AMPA$- and $NMDA$-mediated connections respectively which connect population $x_{kp}^E$ to population $x_{ij}^E$. These inter-WTA coupling weights are fixed in all simulations except the simulations of the perceptual multi-stability model where they are plastic and obey the plasticity rule given in Eq.~\ref{eq:plasticity}. The remaining weights are the intra WTA weights. $\eta^E_{ij}(t)$, $\eta^I_{i}(t)$ are informally treated as derivatives of independent, zero mean, unity variance Wiener processes. We could thus use the Euler-Maruyama method to carry out the noise simulations where in each time step $\Delta t$, the random processes $\eta^E_{ij}(t)$, $\eta^I_{i}(t)$ generate independent Gaussian increments with zero mean and variance $\Delta t$.  $\kappa_i$ is a zero mean, unity variance Gaussian random variable. The term  $\sigma^{OSC}\kappa_i\delta(\phi_i)$ in Eq.~\ref{eq:phase} indicates that at the beginning of each oscillation cycle, $\phi_i=0$, a random increment  $\sigma^{OSC}\kappa_i$ is added to the oscillation phase. This increment is only added once per cycle so the phase has to cross the $2\pi$ point and reset before another increment is added.  $\sigma^{LTU}$ and $\sigma^{OSC}$ are noise scaling factors. $Z_{k\rightarrow i}$ is the phase coupling strength between oscillatory inhibition in WTAs $k$ and $i$. We always use symmetric phase coupling: $Z_{i\rightarrow j} = Z_{j\rightarrow i}$. The phase coupling model follows the Kuramoto model~\cite{Strogatz00}.

The values of all model parameters used in the simulations are given in the supplementary materials. 

\subsection*{Network equivalent MCMC operator}
In MCMC sampling, a stochastic Markovian transition operator $R$ defined over a space ${\bf X}$ is used to generate a stochastic sequence of points, or samples, in {\bf X}: $x_1,x_2,\ldots$, where $x_{n+1}$ is a sample drawn from the conditional probability distribution:
\begin{equation}
p(x_{n+1} = z | x_n=y) = R(y \rightarrow z)
\end{equation}
$R(y \rightarrow z)$ is the probability that the next point in the sequence is $z$ given that the current point is $y$. If $R$ is irreducible (For any two states, $x_1$ and $x_2$, $x_2$ will with non-zero probability appear in the random sequence started from $x_1$ after a finite number of steps), and $R$ is aperiodic (starting from any state $x$, the indices of the positions in the sequence where $x$ has a non-zero probability of occurring do not have a common divisor other than 1) then for any starting point $x_0$, $p(x_n=y)$ approaches the quantity $q(y)$ as $n$ gets large for all $y \in {\bf X}$. In other words, the points $x_n,x_{n+1},\ldots$ are samples drawn from the probability distribution $q(x)$, called the stationary or invariant distribution of the transition operator $R$, for large enough $n$.

Given a network of interacting WTA circuits, our goal is to find a Markovian stochastic transition operator, $T$, that is defined over the space of all network configurations and which approximates the way the network trajectory moves through this configuration space. We here describe how, given a particular network configuration $x_n$, the next configuration $x_{n+1}$ is evaluated. This evaluation procedure implicitly defines the transition operator $T$ and it attempts to approximate the way the configuration of the actual network changes. 
\begin{enumerate}
\item Choose at random a WTA $V$ to be updated. This corresponds to a WTA which has just been released from oscillatory inhibition and is now selecting a winner.
\item Choose at random a subset of the consistency conditions that involve the WTA $V$. Denote this subset as $C$. The WTA circuits involved in this subset are assumed to have the proper phase relations to $V$ that enable them to affect the winner selection in $V$.
\item For each possible state $s_i$ of WTA $V$ (the number of possible states is the number of excitatory populations in the WTA), attach a number $n_i$ that counts the number of consistency conditions in $C$ that are satisfied if $V$ is in state $s_i$. The states of the other WTA circuits are fixed and encoded in $x_n$. Find the maximum of $n_i$ across all states $s_i$ and denote it as $n$. Define $S_C = \{s_k : n_k = n\}$. $S_C$ is thus the set of possible states for WTA $V$ that are favored by the majority of the consistency conditions in $C$. If the current state of $V$ is in $S_C$, then the state of $V$ remains unchanged and $x_{n+1}=x_n$. This reflects the hysteresis mechanism mediated by the slow $NMDA$ current that favors the WTA population that won in the previous inhibition cycle. If the current state of $V$ is not in $S_C$, then a new state for $V$ is selected randomly from $S_C$ and $x_{n+1}$ is updated accordingly. Note that at most one WTA changes its state between $x_n$ and $x_{n+1}$.  
\end{enumerate}

\subsection*{Plasticity rule for inter-WTA connections}
We use the following bi-stable plasticity rule to modulate the inter-WTA connection strengths:
\begin{eqnarray}
\label{eq:plasticity}
\frac{\partial w(t)}{\partial t} = &+& d(w) \\
&+& \eta_{+}(r_{pre},r_{post}) \cdot [r_{pre}(t) -\theta] \cdot \left[r_{post}(t) - \theta \right] \nonumber \\
&-& \eta_{-}(r_{pre},r_{post}) \cdot [r_{pre}(t) -\theta] \cdot [\theta - r_{post}(t)] \nonumber
\end{eqnarray}
where
\[
 d(w) =
  \begin{cases}
   d_{up} & \text{if}\; w(t) > w_{mid} \\
   -d_{down}       & \text{otherwise} 
  \end{cases}
\]
\[
 \eta_+(r_{pre},r_{post})  =
  \begin{cases}
   \eta_{up}  & \text{if}\; r_{pre}(t) > \theta\; \text{and}\; r_{post}(t) > \theta \\
   0      & \text{otherwise} 
  \end{cases}
\]
\[
 \eta_-(r_{pre},r_{post})   =
  \begin{cases}
   \eta_{down}  & \text{if}\; r_{pre}(t) > \theta\; \text{and}\; r_{post}(t) \leqslant \theta \\
   0      & \text{otherwise} 
  \end{cases}
\]

$w(t)$ is the weight. $r_{pre}(t)$ and $r_{post}(t)$ are the firing rates of the source (presynaptic) and target (postsynaptic) populations respectively. This rule is a variation of the Bienenstock-Cooper-Munro (BCM) rule~\cite{Bienenstock_etal82} with hard weight bounds $w_{min}$ and $w_{max}$ (not shown in the equation) and the requirement that $r_{pre}(t)$ has to exceed a threshold, $\theta$, in order to induce any change in the weight $w(t)$. If this requirement is met, potentiation is induced if the postsynaptic activity, $r_{post}(t)$, is above the threshold $\theta$, and depression is induced if the postsynaptic activity is below the threshold. The rates of potentiation and depression induction are controlled by $\eta_{up}$ and $\eta_{down}$ respectively. The rule captures the way potentiation and depression induction depend on the pre- and post-synaptic firing rates~\cite{Sjostrom_etal01}. The rule contains a second component that slowly forces the connection weight to either $w_{min}$ or $w_{max}$ depending on whether the weight is below or above $w_{mid}=\frac{1}{2} (w_{max} + w_{min})$ respectively. The rule is thus bistable. The strength of the bistability drift is controlled by $d_{up}$ and $d_{down}$. Bistable plasticity is computationally less powerful than its counterpart with continuous stable weights \cite{Amit_Fusi92}, but can be argued to be more biologically realistic, due to noise-tolerance and finite synaptic information content.

\subsection*{Details of the perceptual multi-stability model}
To decode the network activity, we first scale the angles of the normalized preferred directions of the  excitatory populations in the hidden WTAs from the $0-180^\circ$ range to the $0-360^\circ$ range. We then perform a vector addition of the modified preferred directions of all active populations of the hidden WTA circuits to obtain the two quantities:
\begin{equation}
\theta = \frac{1}{2} atan2\left( \sum_i \delta_i y_i, \sum_i \delta_i x_i \right)
\label{Eq:decode1}
\end{equation}
\begin{equation}
r= \sqrt{\left(\sum_i \delta_i y_i\right)^2+ \left(\sum_i \delta_i x_i\right)^2}
\label{Eq:decode2}
\end{equation}
where $atan2$ is the two argument quadrant adjusted inverse of the tangent,  $x_i$ ($y_i$) is the x-coordinate (y-coordinate) of the modified preferred direction of population $i$ and $\delta_i \in \{0,1\}$ is the indicator of the activity of population $i$. The summation runs over all excitatory populations in the hidden WTA circuits. The factor $\frac{1}{2}$ in Eq.~\ref{Eq:decode1} puts the decoded angle back in the $0-180^\circ$ range. $r$, the magnitude of the decoded activity vector, can be understood as the confidence of the system in its current angle estimate. When all hidden WTA circuits encode a different angle, $r$ takes a value close to zero, when they all encode the same angle, it takes the maximal value $n_{hid}=6$.

``Training'' the network refers to the following procedure:
 For 30\,seconds, we provide input that clamps the states of all input WTA circuits so that they are all encoding the same orientation, while enabling the global synchronizing oscillation.  Within these 30\,seconds, we cycle through the $n_c=6$ directions each WTA can encode. The configuration of the input WTA circuits is thus forced to represent an unambiguous orientation which would act to influence the hidden WTA circuits so that they are also encoding the same unambiguous orientation. The connections between the hidden WTA circuits change to represent the prior that all hidden WTA circuits usually encode the same orientation.

During testing, we provide inputs that clamp the states of a subset, or of all, input WTA circuits to certain orientations for 4 minutes while withholding the global synchronizing oscillation.  We record the activity of the hidden WTA circuits and decode it into an angle and a magnitude as outlined in Eqs~\ref{Eq:decode1} and~\ref{Eq:decode2}. We interpret the normalized time histogram of this decoded signal as the probability the network assigns to a given $r$, $\theta$ pair.  
\FloatBarrier

\clearpage
\addtocounter{prop}{-2}
\setcounter{figure}{0}
\makeatletter 
\renewcommand{\thefigure}{S\@arabic\c@figure}
\makeatother
\makeatletter 
\renewcommand{\thetable}{S\@arabic\c@table}
\makeatother

\begin{flushleft}
{\Large
\textbf{Rhythmic inhibition allows neural networks to search for maximally consistent states - supplementary material}
}
\\
Hesham Mostafa$^*$, 
Lorenz K. M\"uller, 
Giacomo Indiveri.
\\
Institute for Neuroinformatics, University of Zurich and ETH Zurich, Switzerland \\
E-mail: \{hesham,lorenz,giacomo\}@ini.uzh.ch
\end{flushleft}
\subsection*{Proofs of propositions 1 and 2}
The frequencies of the local oscillatory inhibition in the WTA circuits are different and are not rational multiples of each other. Having one oscillation frequency that is a rational multiple of another is statistically impossible in a physical system with real-valued frequencies. Let ${\bf \Phi}(t) = [\phi_1(t),\phi_2(t),\ldots,\phi_n(t)]$ be the vector of the phases of the oscillatory inhibition in the $n$ WTA circuits in the network at time $t$ where $\phi_i\in[0,2\pi)$. By virtue of the incommensurability of the oscillatory inhibition frequencies, ${\bf \Phi}(t)$ follows a quasi-periodic trajectory.  For any initial phase vector ${\bf \Phi}(0)$, and any $\hat{{\bf \Phi}} \in [0,2\pi]^n$ and $\epsilon > 0$, there is a time $\hat{t}$ such that $|{\bf \Phi}(\hat{t}) - \hat{{\bf \Phi}}| < \epsilon$. The vector of phases ${\bf \Phi}(t)$ thus densely fills up the n-dimensional hyper-torus of all possible phases which implies that ${\bf \Phi}(t)$ cannot have a periodic trajectory. Since it is the phase relations between the WTA circuits that determine the effectiveness of the different consistency conditions, the ergodicity of the phase vector ${\bf \Phi}(t)$ ensures that all possible combinations of the strengths of the consistency conditions are eventually explored. 
In a WTA circuit $W_i$, the oscillatory inhibition is high for $h_i$ seconds and low for $l_i$ seconds during each cycle. In order to prove some useful results about the trajectory of $d(t)$, we require that $h_i > l_j \forall i,j\neq i$. This condition, together with the ergodicity of the phase vector, ${\bf \Phi}$, guarantees that for any WTA circuit $W_r$, there is bound to come an oscillation cycle where the winner selection is unaffected by the activity in any other WTA, because during that cycle, $W_r$ is released from oscillatory inhibition, selects a winner, and gets shut down again by oscillatory inhibition  within $l_r$ seconds, while all other WTA circuits are quiescent. 

\begin{prop}
The only fixed points of $d(t)$ are the configurations that satisfy all consistency conditions
\label{prop:fixedpoint}
\end{prop}
\begin{proof}[Proof]
Assume $d(t)$ has a fixed point $d^*$ which encodes a particular non-changing configuration and that this configuration violates one or more consistency conditions. Let $V_1$ and $V_2$ be two WTA circuits that are part of a violated consistency condition. Due to the ergodicity of the phase vector, ${\bf \Phi}(t)$, there is bound to come an oscillation cycle for $V_2$ which has the following properties: during the part of the cycle when oscillatory inhibition is low in $V_2$, oscillatory inhibition is high in all other WTA circuits except $V_1$ and the peak of activity in the excitatory populations in $V_1$ occurs at the time in which it can dictate the winner in $V_2$. The influence of $V_1$ on $V_2$ is thus unopposed by any other WTA and this influence acts to enforce the violated consistency condition by changing the identity of the winner in $V_2$. Thus, a configuration that violates one or more consistency conditions is unstable. Assume $d(t)$ has reached a point $\bar{d}$ which encodes a configuration that satisfies all consistency conditions. The currently winning population in each WTA circuit will receive from other WTAs either a stronger or an equal input as compared to the other populations in the WTA. In both cases, and due to the hysteresis mechanism mediated by the slow recurrent $NMDA$ current, the winning population in each WTA circuit keeps on winning and the global configuration $\bar{d}$ remains unchanged hence $\bar{d}$ is a fixed point.
\end{proof}

\begin{prop}
\label{prop:aperiodic}
$d(t)$ is not periodic as long as it has not reached a fixed point
\end{prop}
\begin{proof}[Proof]
Since $d(t)$ has not reached a fixed point, $d(t)$ alternates between a number of configurations. Let $V$ be a WTA in which the identity of the winning population changes between these configurations. Assume $d(t)$ has period $T$. Fix $t_0$, at some $t_1$ where $t_0 \le t_1 < t_0+T$, WTA $V$ changes its state, i.e, it selects a winning population that is different from the winning population in the previous cycle. This change is reflected in $d(t)$ at $t_1$ when the oscillatory inhibition goes high in $V$. $V$ has to change its state in the same way at $t_1+nT$ where $n=1,2,\ldots$ by the periodicity assumption of $d(t)$. So $T$ has to be a multiple of the oscillation period of $V$. Since the frequencies of oscillatory inhibition in the WTA circuits are incommensurable, $T$ is incommensurable with the oscillation periods of all WTA circuits except $V$. Hence, there is bound to come an oscillation cycle for $V$ with the following two properties: during the part of the cycle when oscillatory inhibition is low in $V$, oscillatory inhibition is high in all other WTA circuits; and the cycle terminates at $t_1+kT$ when the inhibition in $V$ goes high where $k$ is an integer. So it is impossible for $V$ to select a winner in this cycle that is different from the winner it selected in the previous cycle (in the absence of external influence throughout this cycle, the hysteresis mechanism mediated by the slow $NMDA$ recurrent connections yields the same winner as the previous cycle). This contradicts the initial assumptions about $V$ (that it selects a different winner at $t_1+kT$) and establishes the aperiodicity of $d(t)$.
\end{proof}

\subsection*{Stochastic approximation of the network trajectory}

\begin{figure}
\centering
      \begin{subfigure}[]{0.4\textwidth}
     \includegraphics[width=\textwidth]{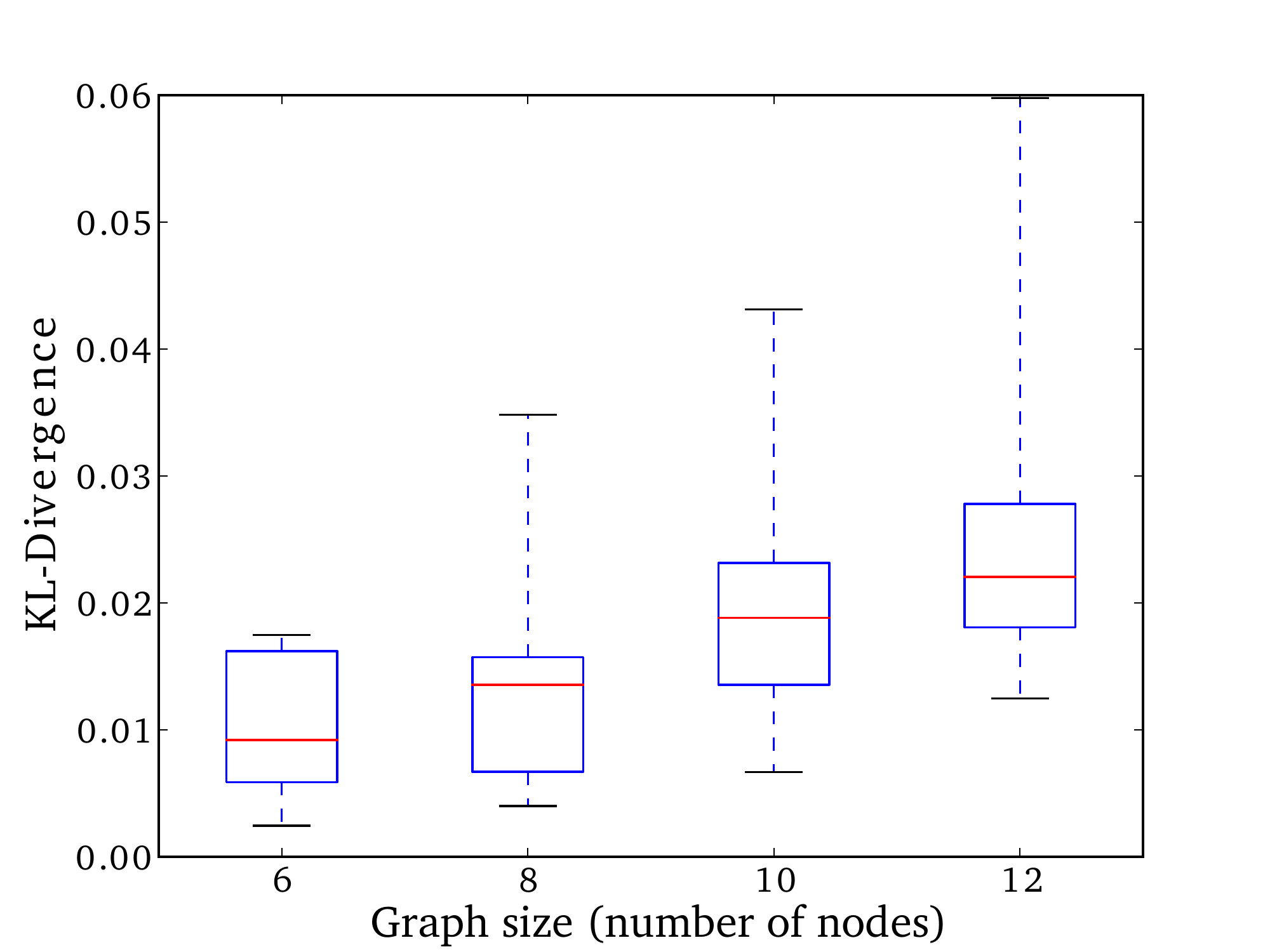} 
     \subcaption{} 
     \label{fig:klnode}
   \end{subfigure}
   \quad
   \quad
   \begin{subfigure}[]{0.4\textwidth}
     \includegraphics[width=\textwidth]{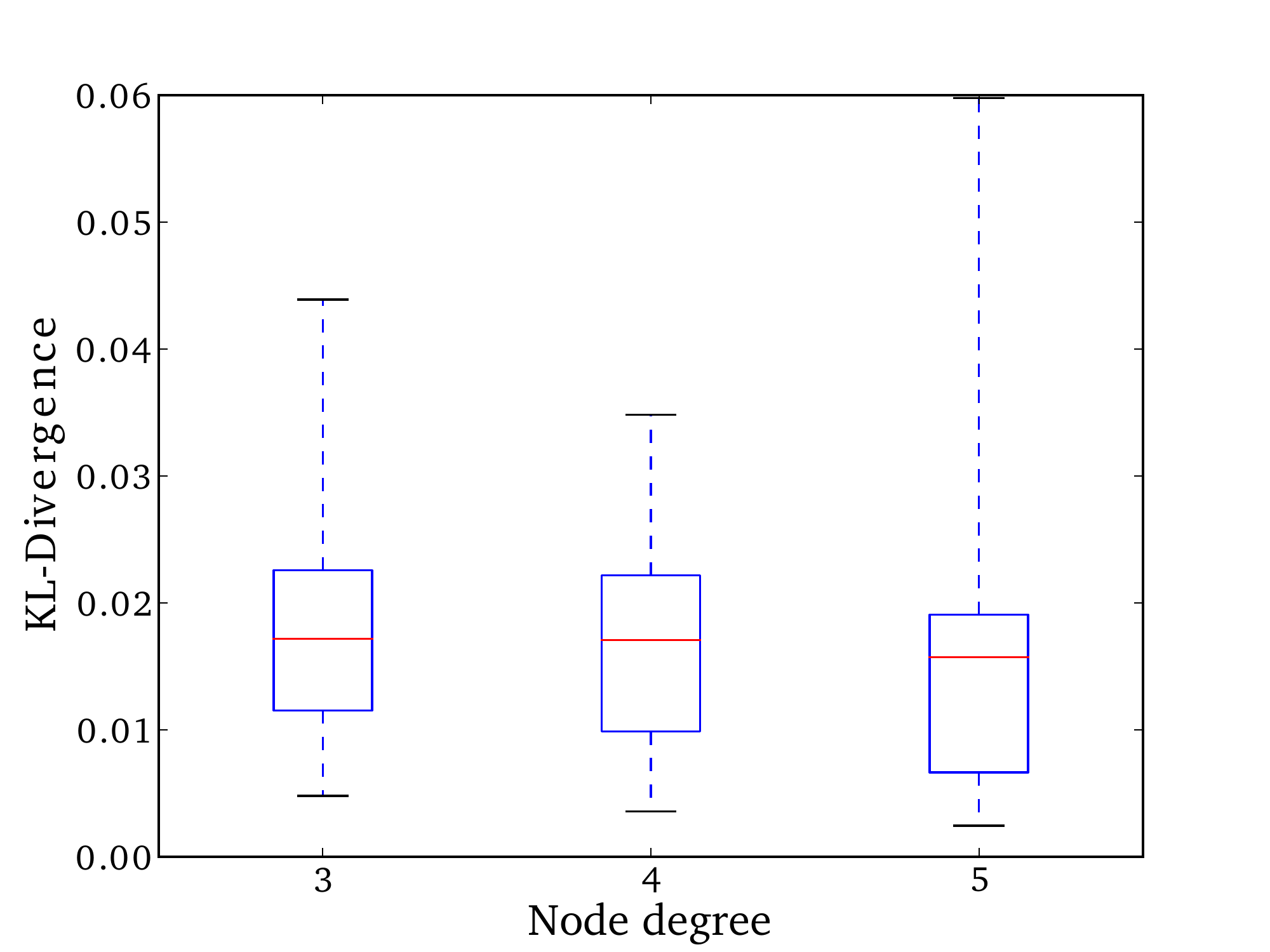}
     \subcaption{} 
     \label{fig:kldeg}
   \end{subfigure}
   \caption{The KL-Divergence between the ``probability distribution'' induced by the network trajectory and the stationary distribution of the equivalent MCMC sampler, calculated for 60 randomly generated regular graphs with sizes (number of nodes) in the range $\{6,8,10,12\}$ and node degrees in the range $\{3,4,5\}$. There are 5 graph instances for each size-degree combination.  In (\subref{fig:klnode}) the KL-Divergence is plotted as a function of the number of nodes in the graph (15 data points per item) and in (\subref{fig:kldeg}), it is plotted as a function of degree (20 data points per item). The red line indicates the median, the box outlines the 1st and 3rd quartile and the whiskers show the full range of the data.}
   \label{fig:KL}
\end{figure}

Based on the topology of the network, an MCMC operator can be constructed that approximates the network trajectory (see Methods section). The stationary probability distribution of this MCMC operator does not exactly reproduce the relative durations of the different configurations in the actual network. There are many reasons for this discrepancy:
\begin{itemize}
\item The state of a WTA (the identity of the winning population) in one oscillation cycle is not only affected by its state in the preceding cycle but also by its state in cycles further in the past. That is because the time constant of the slow recurrent $NMDA$ current is $80\,ms$ and the mean period of the oscillatory inhibition in the WTA circuits is $22\,ms$. When a population wins in one inhibition cycle, its increased recurrent $NMDA$ current takes many cycles to decay which might affect the winner selection process many cycles later. The Markovity property which is fundamental to the construction of $T$ is thus not exact.
\item Some WTA circuits (those that have a higher oscillation frequency) update more often than others. This violates the assumption inherent in step 1 in the MCMC update scheme which is that WTA circuits update equally often on average. In sampling algorithms like Gibbs sampling, differences in the variables update frequencies do not affect the stationary distribution, but in our case it does. One can show that this is a consequence of the fact that the transition operator $T$ does not obey the detailed balance equation  while the transition operator for the Gibbs sampler does.
\item In comparing the statistics of $d(t)$ and those of the sequence generated by $T$, we are comparing the time durations of the different configurations in $d(t)$ to their relative frequencies of occurrence in the MCMC sequence. This way, we are in essence assigning an equal time duration to each MCMC sample which exposes another assumption of the MCMC scheme, namely, that the WTA circuits update their states at equal intervals (the updated state might be the same as the previous one). In the actual network, there is no clear-cut update cycle. The WTA circuits might update their states in very quick succession so the intermediate configurations might have very short duration, or one configuration might persist for a relatively long time before a WTA updates its state (i.e, the inhibition goes high in this WTA and the winner is evaluated). 
\end{itemize}

To further investigate the difference between the network statistics and those of the MCMC operator, we constructed random regular graphs of different sizes and node degrees. From each graph we constructed a network in the following way: each node in the graph is a binary WTA and each edge represents a consistency condition which could either be an equality or inequality condition. We then used Kullback-Leibler(KL) divergence to quantify the difference between the ``probability distribution'' generated by the network and the stationary distribution of the equivalent MCMC operator. The results are shown in Fig~\ref{fig:KL} for different graph/network sizes and node degrees. The above-mentioned points of discrepancy lead to a more pronounced difference in the two sets of statistics in larger networks/graphs. To verify that this is not an effect of finite sample size, we generated a larger number of MCMC samples and ran the network for longer durations; the KL-divergence between the two sets of statistics remained virtually unchanged. The match between the two sets of statistics does not seem to depend on the network/graph degree. 

\subsubsection*{Properties of the network-equivalent MCMC operator} 

The transition operator $T$ for an arbitrary network constructed as outlined above is aperiodic as there is always a non-zero probability that any configuration persists for an arbitrary number of steps in the sequence. Unlike transition operators used in typical MCMC applications, $T$ does not obey detailed balance so the generated Markov chain of samples is irreversible. Also, $T$ is not irreducible in general. 
This means that in general, for networks of interacting WTA circuits, the space of possible configurations can be decomposed into a number of non-communicating sets and one set of transient configurations. Starting the chain from a configuration from the transient set, there is a non-zero probability that this configuration will never be visited again. If started in one of the non-communicating sets, the network trajectory will always stay in this set of configurations. Hence, the long-term frequency of occurrence, or the ``probability distribution'', over the network configurations will in general depend on the initial configuration of the network. 

The reducibility of $T$ is a desirable property in some circumstances. For example, $T$ is reducible for a network where there is a configuration that satisfies all consistency conditions as this configuration persists forever. Having fully consistent configurations as absorbing states enables this kind of networks to be used for solving constraint satisfaction problems. This behavior is useful in the perceptual multi-stability model as it allows the network to form stable percepts if the input is unambiguous. $T$ is also reducible for a network where there is a configuration that violates all consistency conditions as this configuration will never be visited. This could be beneficial as it reduces the space of configurations explored by the network by discarding fully inconsistent states for which there is no evidence, either from external input, or from the consistency conditions in the network. More elaborate networks can be constructed that do not fall into these two categories but that map to a reducible stochastic transition operator by virtue of having two or more non-communicating sets of recurrent states (i.e non-transient states). External input can switch the network configuration from one subset to another. The network can thus explore different parts of the configuration space based on external input. The reducibility of $T$ is, however, undesirable if the goal is to generate samples from a unique probability distribution irrespective of initial conditions which is a typical requirement in many MCMC applications.

\subsection*{High order consistency conditions and the explaining-away effect}
We consider a general consistency condition, $C$, that involves $m$ WTA circuits: $V_1,\ldots,V_m$, where the number of distinct states in WTA $i$ (the number of excitatory populations) is $K_i$. $C$ is simply the set of allowed configurations of the $m$ WTA circuits. We require $C$ to be non-empty. $C$ can be implemented in the following way:
\begin{enumerate}
\item
Define $K = \prod_{i=1}^{i=m-1}K_i$. Introduce a ``hub'' WTA, $H$, which has $K$ excitatory populations in its WTA\@.  
\item For each of the $K$ possible configurations of $V_1,\ldots,V_{m-1}$, connect each of the excitatory populations that are active in this configuration bidirectionally to one distinct excitatory population in $H$. We have thus associated one population in $H$ to each configuration of the WTA circuits $V_1,\ldots,V_{m-1}$.
\item Connect the excitatory populations of WTA $V_m$ to the excitatory populations of $H$ bidirectionally so that the excitatory populations that are active in each of the allowed (consistent) configurations of $C$ connect to an excitatory population in $H$ bidirectionally.
\item Remove all populations in $H$ that are not bidirectionally connected to a population in $V_m$. These populations encode the configurations of the first $m-1$ WTA circuits that are not part of any consistent configurations.
\end{enumerate}
We can show in a way that is analogous to the proofs of propositions 1 and 2 that the described scheme will lead to dynamics that keep flipping the states of WTA circuits $H,V_1,\ldots,V_{m}$ in an aperiodic manner as long as $V_1,\ldots,V_{m}$ are inconsistent according to $C$. The only stable state is one in which $V_1,\ldots,V_{m}$ satisfy $C$ and the state of $H$ reflects the configuration of WTAs $V_1,\ldots,V_{m-1}$.

\begin{figure}
  \centering
   \begin{subfigure}[]{0.6\textwidth}
     \includegraphics[width=\textwidth]{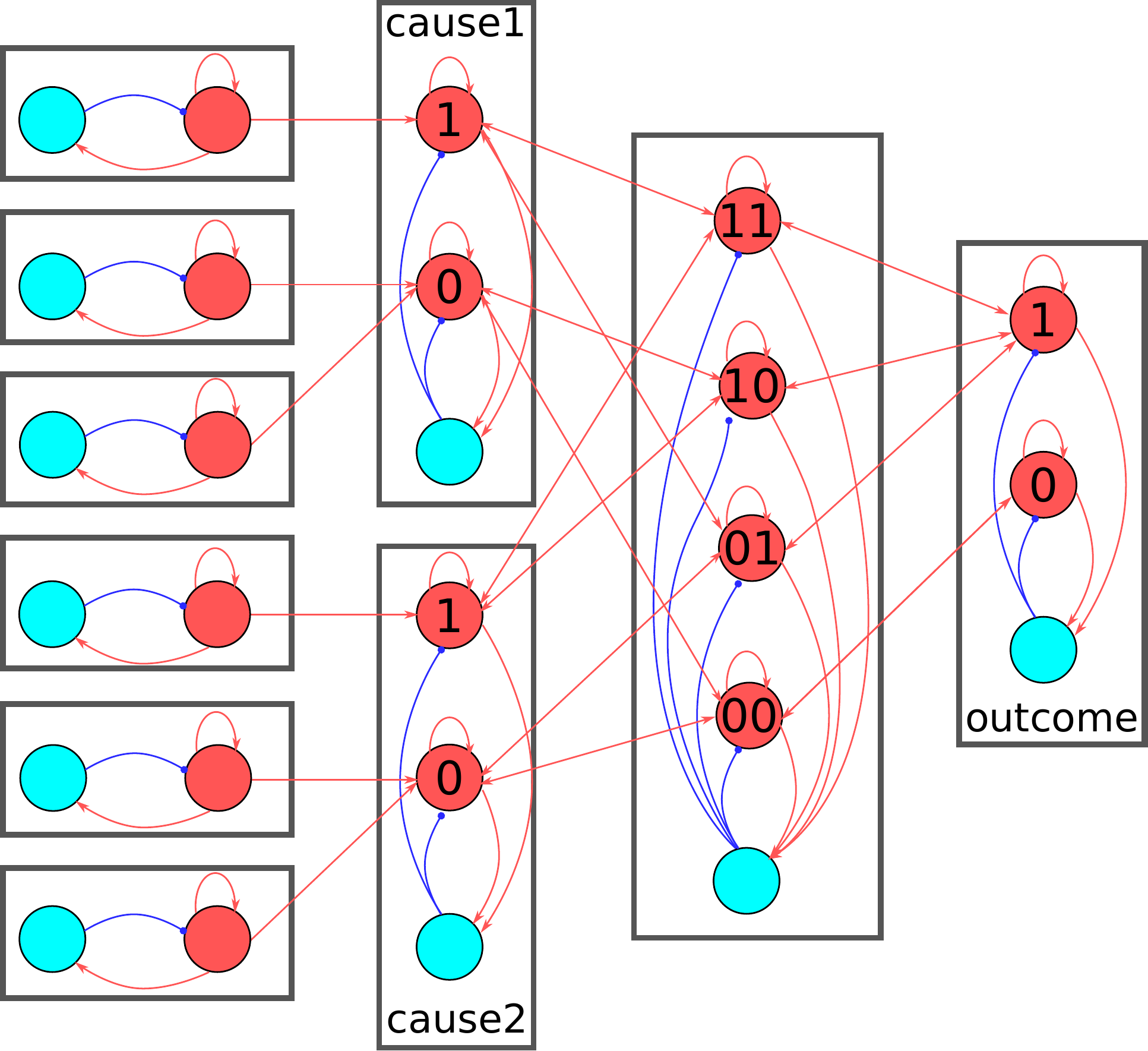} 
     \subcaption{} 
     \label{fig:explainingaway_a}
   \end{subfigure}
   \\
   \begin{subfigure}[]{0.44\textwidth}
     \includegraphics[width=\textwidth]{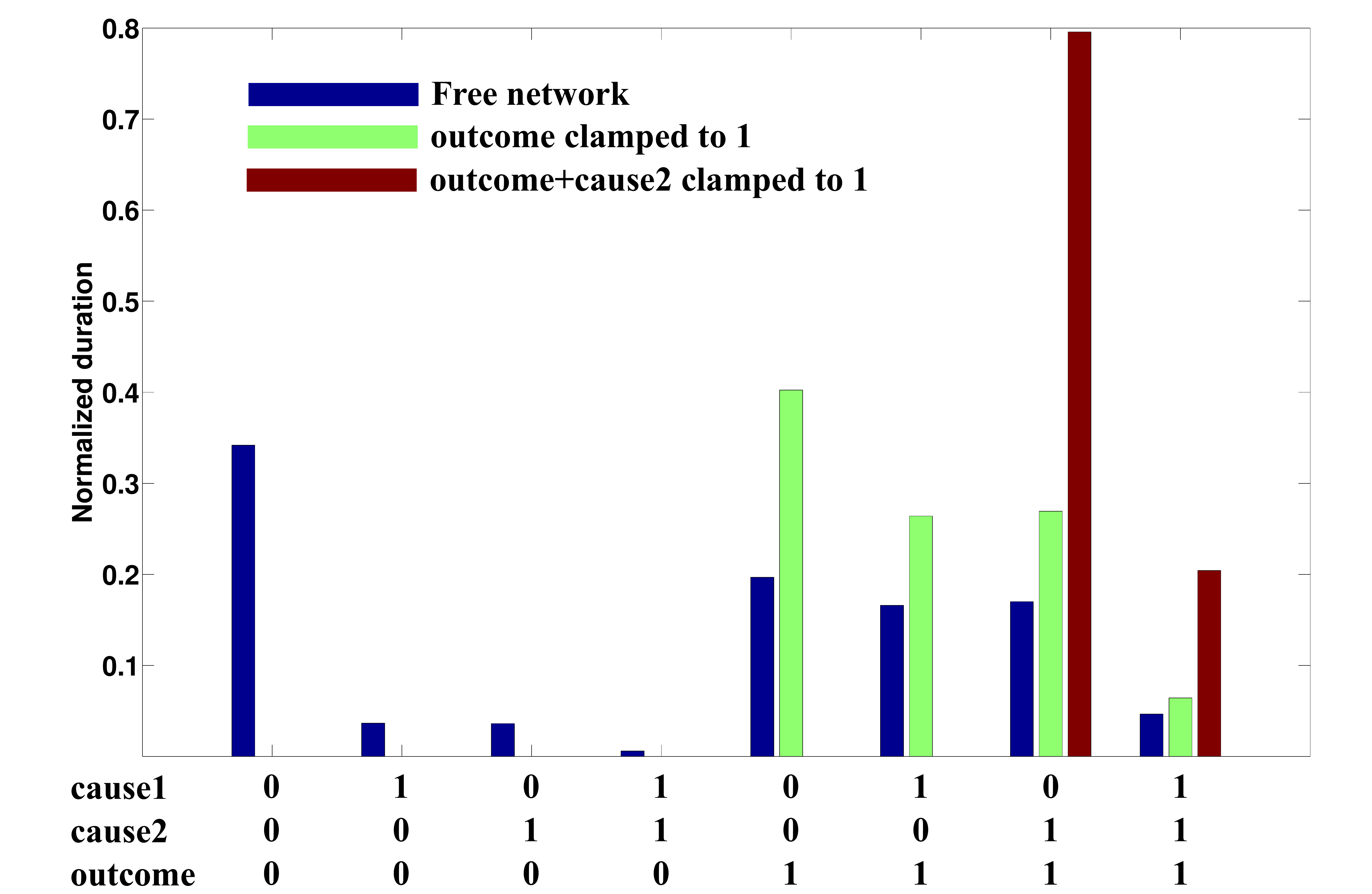} 
     \subcaption{} 
     \label{fig:explainingaway_b}
   \end{subfigure}
   \quad
   \begin{subfigure}[]{0.44\textwidth}
     \includegraphics[width=\textwidth]{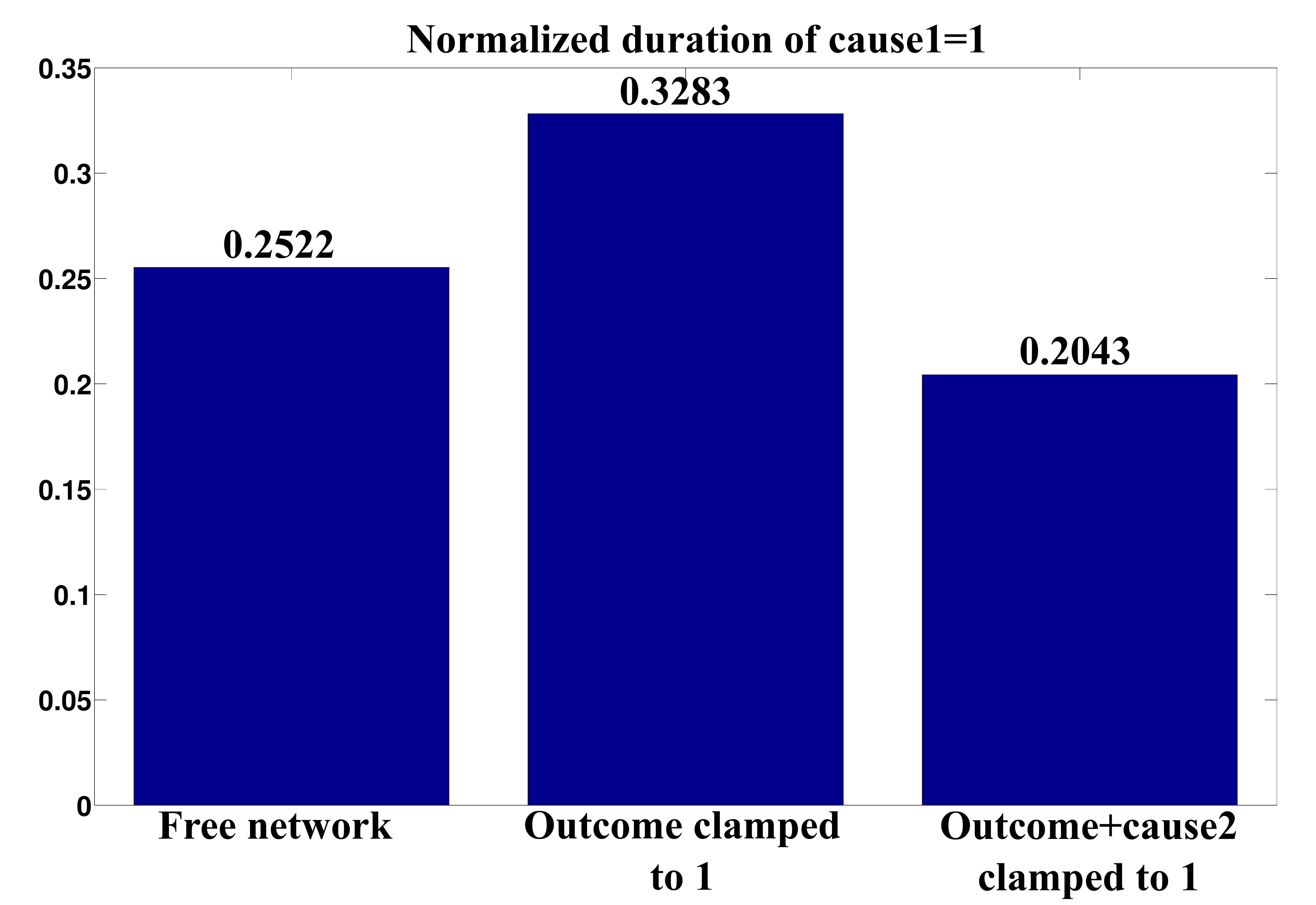} 
     \subcaption{} 
     \label{fig:explainingaway_c}
   \end{subfigure}
\caption{A High-order consistency condition gives rise to the explaining away effect. (\subref{fig:explainingaway_a}) Network topology. The oscillatory inhibition input to each WTA is not shown. The high-order consistency condition encoded in the network connectivity requires the outcome to be an OR function of the two causes. (\subref{fig:explainingaway_b}) Normalized duration of occurrence of the 8 possible states of the \textsf{outcome}, \textsf{cause1}, and \textsf{cause2} WTA circuits under three conditions: network running freely, \textsf{outcome} WTA is clamped to 1, and \textsf{outcome} and \textsf{cause2} clamped to 1. In each condition, the network was simulated for $10^5$\,seconds. (\subref{fig:explainingaway_c}) Normalized duration of the configurations in which \textsf{cause1} is 1 under the three aforementioned conditions. We interpret this as the marginal probability of \textsf{cause1}=1 under the three conditions.}
\label{fig:explainingaway}
\end{figure}
 
High-order consistency conditions can capture complex dependencies between the states of multiple WTA circuits that can not be captured by pairwise connections between the WTA circuits. One example is the dependencies that give rise to the ``explaining away'' phenomenon which is believed to underlie several aspects of visual perception. Explaining away occurs in situations where many causes can give rise to a particular outcome. When this particular outcome is observed, the individual likelihood of each cause increases as at least one of the causes is needed to explain the outcome. If in addition, one of the causes is observed to have taken place, the likelihood of the other causes goes down again. The observed cause explains away the other causes as they are no longer needed to justify the outcome. This form of inter-causal reasoning establishes a dependency between the different causes of a particular outcome when this outcome is observed. Fig.~\ref{fig:explainingaway_a} shows a network used to model the dependencies that give rise to the explaining away effect. Two causes, \textsf{cause1} and \textsf{cause2}, can each give rise to an outcome. The causes and the outcome are binary and they are represented by one WTA each. The causes and the outcome are coupled to a hub WTA so that the binary outcome is effectively an $OR$ function of the two causes. For each cause, we introduce three unitary WTA circuits where two unitary WTA circuits project to the $0$ population and one projects to the $1$ population of the cause WTA\@. That in effect introduces a ``prior'' over the states of the two causes that assigns a higher likelihood to them being $0$. Each WTA  oscillates at a slightly different frequency. The network as a whole can not have a consistent state as each cause can not simultaneously satisfy the contradictory requirements of the unidirectional consistency conditions coming from the unitary WTA circuits. 

Figure~\ref{fig:explainingaway_b} shows the durations of time the network spends in the eight possible configurations of the two causes and the outcome. In the free-running network, there are four configurations that satisfy the high-order consistency condition (the $OR$ relation). Due to the action of the ``prior'', configurations that satisfy the high-order consistency condition have a higher ``probability'' if more of the causes are $0$. The strong ``prior'' can override the high-order consistency condition and assign a high ``probability'' to a configuration in which the two causes are $0$ and which does not satisfy the high-order consistency condition. Similar trends can be seen when the \textsf{outcome} WTA is clamped to $1$ and when both the \textsf{outcome} WTA and the \textsf{cause2} WTA are clamped to $1$. Figure~\ref{fig:explainingaway_c} highlights the explaining away effect. In the free running network, the normalized duration of time in which \textsf{cause1} is equal to $1$ is $0.2522$. This can be interpreted as the ``marginal probability'' of \textsf{cause1}=1. When the outcome is clamped to $1$, this marginal probability increases as one of the causes is needed to explain the outcome (it can now be interpreted as the marginal probability conditioned on \textsf{outcome}=1). When in addition \textsf{cause2} is clamped to $1$, the probability of \textsf{cause1}=1 goes down again as \textsf{cause1} is not needed to explain the outcome when \textsf{cause2} is $1$.

\subsection*{Parameter values used in simulations}
Table~\ref{tab:parameters} contains the parameter values for the network simulations in Figs.1, 3, and 4 in the main text and supplementary Figs.~\ref{fig:KL} and \ref{fig:explainingaway}. The noise and phase coupling terms are only non-zero in the simulations in Fig.~4. Table ~\ref{tab:parametersSud} contains the parameter values used in the Sudoku network in Fig.2. Table~\ref{tab:parameters2} contains the parameter values used in the simulations of the perceptual multi-stability model (Figs.~6 and 7). Parameter values are only shown in tables~\ref{tab:parametersSud} and~\ref{tab:parameters2} if they differ from those in table~\ref{tab:parameters}.

\begin{center}
\begin{table}[ht]
\begin{tabular}{|l|r| l|}
  \hline
  \multicolumn{3}{|c|}{Oscillatory inhibition}
  \\
  \hline
  \hline
  $d$ & 0.6 & Fraction of oscillation cycle with active inhibitory output\\
  $p(f)$ & Uniform(45,46)\,Hz & Distribution of oscillation frequencies\\
  $A$ & 40 Hz & Amplitude of oscillatory inhibition \\
  $w_{osc}^E$ & -3 & Weight oscillatory inhibition to excitatory populations\\
  $w_{osc}^I$ & -1 & Weight oscillatory inhibition to inhibitory populations\\
  \hline \hline 
  \multicolumn{3}{|c|}{Neuron Populations}\\
  \hline \hline
  $\tau^E$ & 0.3 ms & Excitatory population time constant\\
  $\tau^I$ & 0.2 ms & Inhibitory population time constant\\
  $\tau^{NMDA}$ & 80 ms & $NMDA$ time constant\\
  $T^I$ & 8 & Inhibitory population threshold\\
  $T^E$ & -18 & Excitatory population threshold\\
  $w_{rec}^{AMPA}$ & 1.2 & Recurrent intra-WTA non-$NMDA$ excitatory weight \\
  $w_{rec}^{NMDA}$ & 0.004 & Recurrent intra-WTA $NMDA$ excitatory weight \\
  $w_{EI}^{AMPA}$ & 0.5 & Intra-WTA non-$NMDA$ Excitatory to inhibitory weight \\
  $w_{EI}^{NMDA}$ & 0.3 & Intra-WTA $NMDA$ Excitatory to inhibitory weight \\
  $w_{IE}^{GABA}$ & -1 & Intra-WTA Inhibitory to excitatory weight \\
  $w_{k,p\rightarrow i,j}^{AMPA}$ & 0.06 & weight of inter-WTA non-$NMDA$ coupling connection from $x_{k,p}^E$ to $x_{i,j}^E$  \\
  $w_{k,p\rightarrow i,j}^{NMDA}$ & 0.005 & weight of inter-WTA $NMDA$ coupling connection from $x_{k,p}^E$ to $x_{i,j}^E$  \\
  \hline \hline 
  \multicolumn{3}{|c|}{Noise and phase coupling parameters (only for the simulations in Fig.~4 }\\
  \hline \hline
  $\sigma^{LTU}$ & 200 & noise scaling factor in the high noise case\\
  $\sigma^{LTU}$ & 50 & noise scaling factor in the low noise case\\
  $\sigma^{OSC}$ & 0.7 & phase perturbation per cycle for both the low-noise and high-noise case\\
  $Z_{0\rightarrow 3}$,$Z_{3\rightarrow 0}$ & 5 & Phase coupling strength between V0-V3\\  
  \hline 
\end{tabular}
  \caption{Network parameters used for the simulations in Figs.~1, 3, and 4 in the main text and supplementary Figs.~\ref{fig:KL} and \ref{fig:explainingaway} }
  \label{tab:parameters}
\end{table}
\end{center}

\begin{center}
\begin{table}[ht]
\begin{tabular}{|l|r| l|}
  \hline
  \multicolumn{3}{|c|}{Oscillatory inhibition}
  \\
  \hline
  \hline
  $d$ & 0.17 & Fraction of oscillation cycle with active inhibitory output\\
  $p(f)$ & Uniform(40,60)\,Hz & Distribution of oscillation frequencies\\
  $A$ & 40 Hz & Amplitude of oscillatory inhibition \\
  \hline \hline 
  \multicolumn{3}{|c|}{Neuron Populations}\\
  \hline \hline
  $\tau^E$ & 0.5 ms & Excitatory population time constant\\
  $T^I$ & 6 & Inhibitory population threshold\\
  $T^E$ & -2 & Excitatory population threshold\\
  $w_{rec}^{AMPA}$ & 1.8 & Recurrent intra-WTA non-$NMDA$ excitatory weight \\
  $w_{rec}^{NMDA}$ & 0.0001 & Recurrent intra-WTA $NMDA$ excitatory weight \\
  $w_{EI}^{AMPA}$ & 0.6 & Intra-WTA non-$NMDA$ Excitatory to inhibitory weight \\
  $w_{IE}^{GABA}$ & -1.6 & Intra-WTA Inhibitory to excitatory weight \\
  $w_{k,p\rightarrow i,j}^{AMPA}$ & 0.002 & weight of inter-WTA non-$NMDA$ coupling connection from $x_{k,p}^E$ to $x_{i,j}^E$  \\
  $w_{k,p\rightarrow i,j}^{NMDA}$ & 0.0 & weight of inter-WTA $NMDA$ coupling connection from $x_{k,p}^E$ to $x_{i,j}^E$  \\

  \hline 
\end{tabular}
  \caption{Network parameters used for the simulations of the Sudoku network in Fig.~2. Only parameter values that are different from table~\ref{tab:parameters} are shown.}
  \label{tab:parametersSud}
\end{table}
\end{center}

\begin{center}
\begin{table}
\begin{tabular}{|l|r| l|}
  \hline
  \multicolumn{3}{|c|}{Oscillatory inhibition}
  \\
  \hline
  \hline
  $d$ & 0.27 & Fraction of oscillation cycle with active inhibitory output\\
  $f_{G}$ & 45\,Hz& Oscillation frequency to which oscillators couple, when synchronized\\
  $p(f)$ & Uniform(43,53)\,Hz & Distribution of oscillation frequencies\\
  \hline \hline 
  \multicolumn{3}{|c|}{Neuron Populations}\\
  \hline \hline
  $\tau^E$ & 0.5 ms & Excitatory population time constant\\
  $T^E$ & -2 & Excitatory population threshold\\
  $w_{rec}^{NMDA}$ & 0.005 & Recurrent intra-WTA $NMDA$ excitatory weight \\
  $w_{IE}^{GABA}$ & -1.1 & Intra-WTA Inhibitory to excitatory weight \\
  \hline
  \hline 
  \multicolumn{3}{|c|}{Plasticity rule parameters  }\\
  \hline \hline
  $w_{max}$ & 0.03 & Maximal weight of plastic synapse \\
  $w_{min}$ & 0 & Minimal weight of plastic synapse\\
  $\theta$ & 38 Hz & Learning threshold \\
  $\eta_{up}$& $7.5 \times 10^{-5}$/s & potentiation rate \\
  $\eta_{down}$& $3.75 \times10^{-5}$/s & depression rate\\
  $d_{up}$&0.002/s & upward drift rate\\
  $d_{down}$&0.0005/s& downward drift rate\\
  \hline 
  \hline
   \multicolumn{3}{|c|}{Readout WTA (where different from Table~\ref{tab:parameters})}
  \\
  \hline
  \hline
  $w_{in}$ & 0.000216 & Weight from hidden WTA to Readout WTA\\
  $w_{rec}^{AMPA}$ & 1 & Recurrent intra-WTA non-$NMDA$ excitatory weight \\
  $w_{rec}^{NMDA}$ & 0.001 & Recurrent intra-WTA $NMDA$ excitatory weight \\
  $w_{EI}^{AMPA}$ & 0.3 & Intra-WTA non-$NMDA$ Excitatory to inhibitory weight \\

  \hline
\end{tabular}
  \caption{Network parameters used in the perceptual multi-stability model (Figs.~6 and 7). Only parameter values that are different from table~\ref{tab:parameters} are shown.}
  \label{tab:parameters2}
\end{table}
\end{center}

\FloatBarrier


\begin{thebibliography}{10}

\bibitem{Buzsaki_Wang12}
Gy{\"o}rgy Buzs{\'a}ki and Xiao-Jing Wang.
\newblock Mechanisms of gamma oscillations.
\newblock {\em Annual review of neuroscience}, 35:203--225, 2012.

\bibitem{Womelsdorf_etal07}
Thilo Womelsdorf, Jan-Mathijs Schoffelen, Robert Oostenveld, Wolf Singer,
  Robert Desimone, A.K. Engel, and Pascal Fries.
\newblock Modulation of neuronal interactions through neuronal synchronization.
\newblock {\em science}, 316(5831):1609--1612, 2007.

\bibitem{Deco_Rolls03}
G.~Deco and E.~Rolls.
\newblock Attention and working memory: a dynamical model of neuronal activity
  in the prefrontal cortex.
\newblock {\em European Journal of Neuroscience}, 18:2374--2390, 2003.

\bibitem{Tallon-Baudry_etal97}
C.~Tallon-Baudry, O.~Bertrand, C.~Delpuech, and J.~Permier.
\newblock Oscillatory gamma-band (30--70 hz) activity induced by a visual
  search task in humans.
\newblock {\em J. Neurosci.}, 17:722--734, Jan 1997.

\bibitem{Stein_Sarnthein00}
Astrid~Von Stein and Johannes Sarnthein.
\newblock Different frequencies for different scales of cortical integration:
  from local gamma to long range alpha/theta synchronization.
\newblock {\em International Journal of Psychophysiology}, 38(3):301--313,
  2000.

\bibitem{buzsaki_draguhn04}
Gy{\"o}rgy Buzs{\'a}ki and Andreas Draguhn.
\newblock Neuronal oscillations in cortical networks.
\newblock {\em Science}, 304(5679):1926--1929, 2004.

\bibitem{Cardin_etal09}
J.A. Cardin, Marie Carl{\'e}n, Konstantinos Meletis, Ulf Knoblich, Feng Zhang,
  Karl Deisseroth, Li-Huei Tsai, and C.I. Moore.
\newblock Driving fast-spiking cells induces gamma rhythm and controls sensory
  responses.
\newblock {\em Nature}, 459(7247):663--667, 2009.

\bibitem{Jadi_Sejnowski14}
M.P. Jadi and T.J. Sejnowski.
\newblock Regulating cortical oscillations in an inhibition-stabilized network.
\newblock {\em Proceedings of the IEEE}, 102(5):830--842, May 2014.

\bibitem{Brunel_Wang03}
N.~Brunel and X.~J. Wang.
\newblock What determines the frequency of fast network oscillations with
  irregular neural discharges? {I}. synaptic dynamics and excitation-inhibition
  balance.
\newblock {\em Journal of Neurophysiology}, 90:415--430, 2003.

\bibitem{Douglas_Martin04}
R.J. Douglas and K.A.C. Martin.
\newblock Neural circuits of the neocortex.
\newblock {\em Annual Review of Neuroscience}, 27:419--51, 2004.

\bibitem{Bosman_etal12}
C.A. Bosman, Jan-Mathijs Schoffelen, Nicolas Brunet, Robert Oostenveld, A.M.
  Bastos, Thilo Womelsdorf, Birthe Rubehn, Thomas Stieglitz, Peter {De Weerd},
  and Pascal Fries.
\newblock Attentional stimulus selection through selective synchronization
  between monkey visual areas.
\newblock {\em Neuron}, 75(5):875--888, 2012.

\bibitem{Fries05}
Pascal Fries.
\newblock A mechanism for cognitive dynamics: neuronal communication through
  neuronal coherence.
\newblock {\em Trends in cognitive sciences}, 9(10):474--480, 2005.

\bibitem{von-Helmholtz_Southall25}
H.~von Helmholtz and J.P.C Southall.
\newblock {\em Treatise on physiological optics. III. The perceptions of
  vision.}
\newblock New York, 1925.

\bibitem{Friston03}
Karl Friston.
\newblock Learning and inference in the brain.
\newblock {\em Neural Networks}, 16(9):1325--1352, 2003.

\bibitem{Berkes_etal11}
Pietro Berkes, Gerg{\H{o}} Orb{\'a}n, M{\'a}t{\'e} Lengyel, and J{\'o}zsef
  Fiser.
\newblock Spontaneous cortical activity reveals hallmarks of an optimal
  internal model of the environment.
\newblock {\em Science}, 331(6013):83--87, 2011.

\bibitem{hopfield_tank85}
J.J. Hopfield and D.W. Tank.
\newblock “neural” computation of decisions in optimization problems.
\newblock {\em Biological cybernetics}, 52(3):141--152, 1985.

\bibitem{Hopfield08}
J.J. Hopfield.
\newblock Searching for memories, sudoku, implicit check bits, and the
  iterative use of not-always-correct rapid neural computation.
\newblock {\em Neural computation}, 20(5):1119--1164, 2008.

\bibitem{Mostafa_etal13b}
H.~Mostafa, L.~K. M\"uller, and G.~Indiveri.
\newblock Recurrent networks of coupled winner-take-all oscillators for solving
  constraint satisfaction problems.
\newblock In C.J.C. Burges, L.~Bottou, M.~Welling, Z.~Ghahramani, and K.Q.
  Weinberger, editors, {\em Advances in Neural Information Processing Systems
  ({NIPS})}, volume~26, pages 719--727, 2013.

\bibitem{Buesing_etal11}
Lars Buesing, Johannes Bill, Bernhard Nessler, and Wolfgang Maass.
\newblock Neural dynamics as sampling: A model for stochastic computation in
  recurrent networks of spiking neurons.
\newblock {\em PLoS computational biology}, 7(11):e1002211, 2011.

\bibitem{Pecevski_etal11}
Dejan Pecevski, Lars Buesing, and Wolfgang Maass.
\newblock Probabilistic inference in general graphical models through sampling
  in stochastic networks of spiking neurons.
\newblock {\em PLoS computational biology}, 7(12):e1002294, 2011.

\bibitem{habenschuss_etal13}
Stefan Habenschuss, Zeno Jonke, and Wolfgang Maass.
\newblock Stochastic computations in cortical microcircuit models.
\newblock {\em PLoS computational biology}, 9(11):e1003311, 2013.

\bibitem{ainsworth_etal11}
Matthew Ainsworth, Shane Lee, M.O. Cunningham, A.K. Roopun, R.D. Traub, and
  N.J. Kopelland~M.A. Whittington.
\newblock Dual gamma rhythm generators control interlaminar synchrony in
  auditory cortex.
\newblock {\em The Journal of Neuroscience}, 31(47):17040--17051, 2011.

\bibitem{Fries09}
Pascal Fries.
\newblock Neuronal gamma-band synchronization as a fundamental process in
  cortical computation.
\newblock {\em Annual review of neuroscience}, 32:209--224, 2009.

\bibitem{Sohal_etal09}
V.~S. Sohal, F.~Zhang, O.~Yizhar, and K.~Deisseroth.
\newblock Parvalbumin neurons and gamma rhythms enhance cortical circuit
  performance.
\newblock {\em Nature}, 459(7247):698--702, 2009.

\bibitem{Markram_etal04}
H.~Markram, M.~Toledo-Rodriguez, Y.~Wang, A.~Gupta, G.~Silberberg, and C.~Wu.
\newblock Interneurons of the neocortical inhibitory system.
\newblock {\em Nature Reviews Neuroscience}, 5(10):793--807, 2004.

\bibitem{Ray_Maunsell10}
Supratim Ray and J.H.R Maunsell.
\newblock Differences in gamma frequencies across visual cortex restrict their
  possible use in computation.
\newblock {\em Neuron}, 67(5):885--896, 2010.

\bibitem{Burns_etal10}
S.P. Burns, P.~Samuel, D.~Xing, M.J. Shelley, and R.M. Shapley.
\newblock Searching for autocoherence in the cortical network with a
  time-frequency analysis of the local field potential.
\newblock {\em The Journal of Neuroscience}, 30(11):4033--4047, 2010.

\bibitem{Amit_Brunel97}
D.J. Amit and N.~Brunel.
\newblock Model of global spontaneous activity and local structured activity
  during delay periods in the cerebral cortex.
\newblock {\em Cerebral Cortex}, 7:237--252, 1997.

\bibitem{Knight00}
Bruce~W Knight.
\newblock Dynamics of encoding in neuron populations: some general mathematical
  features.
\newblock {\em Neural Computation}, 12(3):473--518, 2000.

\bibitem{Mattia_Del-Giudice02}
Maurizio Mattia and Paolo {Del Giudice}.
\newblock Population dynamics of interacting spiking neurons.
\newblock {\em Physical Review E}, 66(5):051917, 2002.

\bibitem{Friston94}
K.J. Friston.
\newblock {Functional and effective connectivity in neuroimaging: a synthesis}.
\newblock {\em Human Brain Mapping}, 2(1-2):56--78, 1994.

\bibitem{Massimini_etal05}
Marcello Massimini, Fabio Ferrarelli, Reto Huber, S.K. Esser, Harpreet Singh,
  and Giulio Tononi.
\newblock Breakdown of cortical effective connectivity during sleep.
\newblock {\em Science}, 309(5744):2228--2232, 2005.

\bibitem{Strogatz00}
S.H. Strogatz.
\newblock From kuramoto to crawford: exploring the onset of synchronization in
  populations of coupled oscillators.
\newblock {\em Physica D: Nonlinear Phenomena}, 143(1):1--20, 2000.

\bibitem{Ackley_etal85}
David~H Ackley, Geoffrey~E Hinton, and Terrence~J Sejnowski.
\newblock A learning algorithm for boltzmann machines.
\newblock {\em Cognitive science}, 9(1):147--169, 1985.

\bibitem{Axmacher_etal06}
Nikolai Axmacher, Florian Mormann, Guillen Fern{\'a}ndez, C.E. Elger, and
  Juergen Fell.
\newblock Memory formation by neuronal synchronization.
\newblock {\em Brain research reviews}, 52(1):170--182, 2006.

\bibitem{Mamassian_Goutcher05}
Pascal Mamassian and Ross Goutcher.
\newblock Temporal dynamics in bistable perception.
\newblock {\em Journal of Vision}, 5(4):7, 2005.

\bibitem{Gershman_etal12}
Samuel~J Gershman, Edward Vul, and Joshua~B Tenenbaum.
\newblock Multistability and perceptual inference.
\newblock {\em Neural computation}, 24(1):1--24, 2012.

\bibitem{Wong_etal07}
K.-F. Wong, A.C. Huk, M.N. Shadlen, and X.-J. Wang.
\newblock Neural circuit dynamics underlying accumulation of time-varying
  evidence during perceptual decision making.
\newblock {\em Frontiers in Computational Neuroscience}, 1:6--, 2007.

\bibitem{Rumelhart_McClelland86}
D.E. Rumelhart and J.L. McClelland.
\newblock {\em Parallel distributed processing: explorations in the
  microstructure of cognition. Volume 1. Foundations}.
\newblock MIT Press, Cambridge, MA, USA, 1986.

\bibitem{Weiss_Rappelsberger00}
Sabine Weiss and Peter Rappelsberger.
\newblock Long-range eeg synchronization during word encoding correlates with
  successful memory performance.
\newblock {\em Cognitive Brain Research}, 9(3):299--312, 2000.

\bibitem{Fell_Axmacher11}
Juergen Fell and Nikolai Axmacher.
\newblock The role of phase synchronization in memory processes.
\newblock {\em Nature Reviews Neuroscience}, 12(2):105--118, 2011.

\bibitem{Miltner_etal99}
W.H.R. Miltner, Christoph Braun, Matthias Arnold, Herbert Witte, and Edward
  Taub.
\newblock Coherence of gamma-band eeg activity as a basis for associative
  learning.
\newblock {\em Nature}, 397(6718):434--436, 1999.

\bibitem{Markram_etal97}
H.~Markram, J.~L{\"u}bke, M.~Frotscher, and B.~Sakmann.
\newblock Regulation of synaptic efficacy by coincidence of postsynaptic {AP}s
  and {EPSP}s.
\newblock {\em Science}, 275:213--215, 1997.

\bibitem{Bienenstock_etal82}
E.L. Bienenstock, L.N. Cooper, and P.W. Munro.
\newblock {Theory for the development of neuron selectivity: orientation
  specificity and binocular interaction in visual cortex}.
\newblock {\em Jour. Neurosci.}, 2(1):32--48, 1982.

\bibitem{Sjostrom_etal01}
P.J. Sj\"{o}str\"{o}m, G.G. Turrigiano, and S.B. Nelson.
\newblock {Rate, Timing, and Cooperativity Jointly Determine Cortical Synaptic
  Plasticity}.
\newblock {\em Neuron}, 32(6):1149--1164, December 2001.

\bibitem{Amit_Fusi92}
D.J. Amit and S.~Fusi.
\newblock Constraints on learning in dynamic synapses.
\newblock {\em Network: Computation in Neural Systems}, 3(4):443--464, 1992.

\end{thebibliography}
\end{document}